\documentclass[twocolumn, pra]{revtex4-1}

\usepackage{amsmath}
\usepackage{bm}
\usepackage{graphicx}
\usepackage{mathrsfs}
\usepackage{amsfonts}
\usepackage{amssymb}
\usepackage{amsthm}
\usepackage{algorithm}
\usepackage{algorithmic}
\usepackage{enumitem}

	\newtheorem{defi}{Definition}

    \newtheorem{result}{Result}

\usepackage{comment}
\usepackage[usenames,dvipsnames]{color}
\usepackage[colorlinks=true, citecolor=blue,urlcolor=blue, linkcolor=blue]{hyperref}

\usepackage{soul}

\soulregister\ref{7}
\soulregister\cite{7}

\graphicspath{{../old_files/Paper1v7/}}

\begin{document}

\title{Hamiltonian identifiability assisted by single-probe measurement}

\author{Akira Sone}
\author{Paola Cappellaro}
\email{pcappell@mit.edu}
\affiliation{Research Laboratory of Electronics and Department of Nuclear Science and Engineering, Massachusetts Institute of Technology, Cambridge, MA 02139}
\begin{abstract}
We study the Hamiltonian identifiability of a many-body spin-$1/2$ system assisted by the measurement on a single quantum probe based on the eigensystem realization algorithm approach employed in  
Phys. Rev. Lett. \textbf{113}, 080401 (2014). We demonstrate a potential application of Gr\"obner basis to the identifiability test of the Hamiltonian, and provide the necessary experimental resources, such as the lower bound in the number of the required sampling points, the upper bound in total required evolution time, and thus the total measurement time. Focusing on the examples of the identifiability in the spin chain model with nearest-neighbor interaction, we classify the spin-chain Hamiltonian based on its identifiability, and provide the control protocols to engineer the non-identifiable Hamiltonian to be an identifiable Hamiltonian.  
\end{abstract}

\maketitle

\section{Introduction}
\label{sec:introduction}

Quantum system identification is a prerequisite for any technology in quantum engineering, in order to build reliable devices for quantum computation, quantum cryptography or quantum metrology. 
The dynamics of a closed quantum system is dictated by its Hamiltonian; therefore, \textit{Hamiltonian identification} is a central problem. 
In particular, characterizing  many-body qubit Hamiltonians is essential in the quest of building a scalable quantum information processor. 
The development of  system identification techniques is expected to have impact in diverse fields, such as  structural determination of a complex molecule~\cite{Ajoy16x, Ajoy15, Lovchinsky16}, biosensing~\cite{Cooper14, Barry16} and studying magnetism at the nanoscale  \cite{Sar14, Wolfe16}. 

Various methodologies have been developed for this task, including quantum process tomography~\cite{Burgarth09a, Burgarth09b, Franco09, Wang16}, Bayesian analysis~\cite{Granade12, Schirmer15, Sergeevich11}, compressive sensing~\cite{Shabini11, Magesan13c, Arai15}, and eigensystem realization algorithm~\cite{Zhang14, Zhang15, Hou13}.
Not only are many of these techniques quite complex, but they also often assume complete access to the system to be identified: full controllability and observability via the coupling of the target quantum system with a classical apparatus. As this is difficult in practice,  we consider  performing quantum system identification using the coupling of the target system with a \textit{quantum probe}~\cite{Burgarth09a,Burgarth09b, Franco09, Zhang14, Zhang15}.  

Recent  progress in quantum metrology assisted by single quantum probe has demonstrated the ability to achieve precise estimation of a few unknown parameters~\cite{Maletinsky12, Boss16}. These advances now open experimental opportunities for multiple parameter estimation, while offering the advantage of nanoscale probing and coherent coupling of complex quantum systems.

Classical linear system identification has been a widely studied subject for the past decades~\cite{Ljung99}. A popular system identification method for the linear time-invariant (LTI) systems is the eigensystem realization algorithm (ERA)~\cite{Katayama06}. 
ERA has been applied in several fields to study  classical systems, from structural engineering~\cite{Caicedo04} to aerospace engineering~\cite{Moncayo10}. The first applications of ERA to quantum system identification both for close  and open systems were given by Zhang and Sarvoar~\cite{Zhang14, Zhang15}, and a robust estimation was experimentally demonstrated for a closed quantum system~\cite{Hou13}. In this paper, we employ ERA to analyze the required experimental resources to achieve Hamiltonian identification. 
To achieve this, we propose a systematic algorithm to test Hamiltonian  identifiability by employing the idea of Gr\"obner basis, which is an essential concept in the commutative algebra and algebraic geometry~\cite{Cox15, Cox04, Froeberg97, Becker93}.
In particular, we use these techniques to explore what  Hamiltonian models can be identified when restricting our access to a single quantum probe. Further, we provide a lower bound in the number of sampling points required to fully identify the Hamiltonian, which sets an upper bound for the total evolution time and thus the total measurement time.\\
The paper is structured as follows. In Sec.~\ref{sec:preliminary}, we give a brief review of ERA and the Gr\"obner basis, with further details in the Appendixes. In Sec.~\ref{sec:test}, we define the identifiability of many-body spin-1/2 Hamiltonians. We also propose a systematic algorithm to test the identifiability of the Hamiltonian by employing Gr\"obner basis. These results lead us to derive, in Sec.~\ref{sec:boundpoint}, bounds on the resources required for Hamiltonian identification. In Sec.~\ref{sec:examples}, we show some examples of the Hamiltonian identifiability test in the spin chain system by focusing on four spin models. For the identifiable Hamiltonians, we also clarify the relation between the dimension of the spin chain and the experimental resources. In Sec.~\ref{sec:control}, we discuss the application of the external control to achieve the identifiability transfer based on average Hamiltonian theory. Finally, in Sec.~\ref{sec:robustness} we assess the estimation performance of ERA for Hamiltonian identification in the presence of noise, before presenting our conclusions in Sec.~\ref{sec:conclusion}.

\section{Preliminary}
\label{sec:preliminary}
\subsection{Eigensystem Realization Algorithm} 
\label{sec: ERA}
The eigensystem realization algorithm  allows one to obtain a new \textit{realization} of a system from the experimental data, from which a transfer function is derived. The parameters can then be extracted by solving a system of polynomial equations derived from equalizing the new realization transfer function with the transfer function obtained from the state-space representation of the system.  
 Let us review the ERA approach introduced in~\cite{Zhang14} in the context of  Hamiltonian identification assisted by single-probe measurement. 
The Hamiltonian $H$ can be generally parameterized as:
\begin{equation}
H=\sum_{m=1}^{M}\theta_m S_m,
\label{eq:Hamiltonian1}
\end{equation}
where $\theta_m\in\mathbb{R}\setminus\{0\}$ are the unknown non-zero parameters to be determined, and $S_m$ are Hermitian operators. 
For an interacting $N$ spin-1/2 system, $iS_m$'s are the independent elements of $SU(2^{N})$. 
Let us define a set $G_0=\{O_i\}$, which we call \textit{observable set}, 
of operators that we can directly measure and such that $[O_i,H]\neq0$. In our scenario, we will typically consider only observables $O_1$ on the first spin, which is our quantum probe. 
 Let $\Gamma$ be the set of operators constructing the Hamiltonian, i.e. $\Gamma=\{S_m|iS_m\in SU(2^{N}), m=1,2,\cdots, M\}$. 
Then, an iterative procedure $G_j\equiv G_{j-1}\cup [G_{j-1},\Gamma]$, with $[G_{j-1},\Gamma]\equiv \{O_{i}|\text{tr}(O_{i}^{\dagger}[\eta,\gamma])\neq 0, \eta\in G_{j-1},\gamma\in \Gamma\}$, generates a set $G$ of  dimension $n \le4^{N}-1$,
\begin{align*}
G=\{O_k|iO_k\in SU(2^{N}),k=1,2,\cdots,n\},
\end{align*}
 called the \textit{accessible set}. $G$ describes all the operators that become \textit{indirectly observable} when measuring the single quantum probe, thanks to the dynamics of the system. In particular, $G$ typically includes spin correlations. Let $\rho_0$ be the initial state of the system, so that the expectation value of $O_k$ is given by $x_k(t)\equiv\langle O_k(t)\rangle=\text{tr}[\rho_0 O_k(t))]$. Then, the expectation values of the accessible set elements form the \textit{coherent} vector $\mathbf{x}(t)=(x_1(t),\cdots, x_n(t))^{T}\in\mathbb{R}^{n}$ with time evolution 
\begin{align*}
\dot{\mathbf{x}}(t)=\tilde{\mathbf{A}}\mathbf{x}(t),
\end{align*}
where $\tilde{\mathbf{A}}\in\mathbb{R}^{n\times n}$ is a skew-symmetric matrix, which contains the parameters $\theta_m$ as its off-diagonal elements. Generally, $\tilde{\mathbf{A}}$ does not necessarily depend on \textit{all} the parameters. Only when the dynamics correlates all the spins to the quantum probe, $\tilde{\mathbf{A}}$ contains all the parameters, which is a necessary condition for  system identification. Let $y(t)\in\mathbb{R}$ be the output data obtained by the output matrix $\mathbf{C}\in\mathbb{R}^{n}$. In our model, the shape of $\mathbf{C}$ is restricted because we only consider the measurement on the quantum probe. Then, we can obtain the following state-space representation:
\begin{equation}
\begin{split}
\dot{\mathbf{x}}(t)&=\tilde{\mathbf{A}}\mathbf{x}(t)\\
y(t)&=\mathbf{C}\mathbf{x}(t).
\end{split}
\label{eq:linear1}
\end{equation}
It is useful to define the corresponding discrete-time representation because the output data will be only acquired at the discrete-time steps:
\begin{align*}
\begin{split}
\mathbf{x}(j+1)&=\mathbf{A}\mathbf{x}(j)\\
y(j)&=\mathbf{C}\mathbf{x}(j),
\end{split}
\end{align*}   
where we set $\mathbf{x}(j)\equiv x(j\Delta t)$, $y(j)\equiv y(j\Delta t)$ and $\mathbf{A}\equiv e^{\tilde{\mathbf{A}}\Delta t}$. Note that since any matrix exponential is nonsingular, we have:
\begin{equation}
\text{rank}(\mathbf{A})=n,
\label{eq:rankA}
\end{equation}
{where $n$ is called \textit{model order}~\cite{Ljung99}.}
From Eq.~(\ref{eq:linear1}), we can obtain the  transfer function $T(s)=\mathbf{C}(sI_n-\tilde{\mathbf{A}})^{-1}\mathbf{x}(0)$, and $[\tilde{\mathbf{A}},\mathbf{C},\mathbf{x}(0)]$ is called the \textit{realization} of $T(s)$. The coefficients of the Laplace variable $s$ in both numerator and denominator of $T(s)$ are polynomials of the parameters $\theta_m$.

In order to perform ERA, we construct a Hankel matrix and shifted Hankel matrix with the output data as their elements:
\begin{equation}
\mathbf{H}_{rs}(0)=
\begin{pmatrix}
y(0)&y(1)&\cdots& y(s-1)\\
y(1)&y(2)&\cdots& y(s)\\
\vdots&\vdots&\ddots&\vdots\\
y(r-1)&y(r)&\cdots& y(r+s-2)
\end{pmatrix}
\label{eq:Hankel0}
\end{equation} 
\begin{equation}
\mathbf{H}_{rs}(1)=
\begin{pmatrix}
y(1)&y(2)&\cdots& y(s)\\
y(2)&y(3)&\cdots& y(s+1)\\
\vdots&\vdots&\ddots&\vdots\\
y(r)&y(r+1)&\cdots& y(r+s-1)
\end{pmatrix},
\label{eq:Hankel01}
\end{equation}
where {$r$ and $s$ must satisfy $r,s\ge n$, which is the necessary condition for ERA (see Appendix~\ref{sec:Hankel} for details). }
From the singular value decomposition (SVD) of $\mathbf{H}_{rs}(0)$ and the expression of $\mathbf{H}_{rs}(1)$, we can obtain a new realization $[\tilde{\mathbf{A}}_{\text{est}},\mathbf{C}_{\text{est}},\mathbf{x}_{\text{est}}(0)]$ and thus a new corresponding transfer function $T_{\text{est}}(s)=\mathbf{C}_{\text{est}}(sI_n-\tilde{\mathbf{A}}_{\text{est}})^{-1}\mathbf{x}_{\text{est}}(0)$ (see Appendix~\ref{sec:Hankel} for details). Since $T(s)$ and $T_{\text{est}}(s)$ describe the same system, we must have:
\begin{equation}
T(s)= T_{\text{est}}(s).
\label{eq:identity}
\end{equation}
Therefore, the parameters can be found by solving the system of polynomial equations derived from Eq.~(\ref{eq:identity}). We thus reduce the problem of Hamiltonian identifiability to the question of solvability of a system of polynomial equations. 

\subsection{Gr\"obner basis}
The Gr\"obner basis, first introduced by Buchberger in ~\cite{Buchberger06},  is a {systematic method} to solve a system of multivariate polynomial equations and determining its solvability over the complex field $\mathbb{C}$. Following~\cite{Cox15, Cox04, Froeberg97, Becker93},
let us denote by $\mathbb{C}[\theta_1,\cdots,\theta_M]$  the polynomial ring. Suppose that from Eq.~(\ref{eq:identity}), we have obtained the following system of polynomial equations: 
\begin{equation}
f_1(\theta_1,\cdots,\theta_M)=\cdots=
f_p(\theta_1,\cdots,\theta_M)=0.
\label{eq:systemeq1}
\end{equation}
$f_1,\cdots, f_p$ generates a polynomial ideal $\mathcal{I}=\langle f_1,\cdots,f_p\rangle$ with radical $\sqrt{\mathcal{I}}$ (see Appendix~\ref{sec:Groebner} for details). When $\mathcal{I}=\sqrt{\mathcal{I}}$, the ideal is called a radical ideal. 

Fixing a monomial ordering for polynomials $f\in\mathbb{C}[\theta_1,\cdots,\theta_M]$, such as the lexicographic ordering,  we denote by $\text{LM}(f)$ and $\text{LT}(f)$  the leading monomials and leading terms of the polynomial $f$, respectively. 
From the Hilbert basis theorem, there exists a finite set $\mathcal{G}(\mathcal{I})=\{g_1,\cdots, g_t\}$, such that $\mathcal{I}=\langle\mathcal{G}(\mathcal{I})\rangle=\langle g_1,\cdots,g_t\rangle$, where for every polynomial $f\in\mathcal{I}\setminus\{0\}$, $\text{LT}(f)$ is divisible by $\text{LT}(g_j)$ for some $j$. 
Here, $\mathcal{G}$ is called a \textit{Gr\"obner basis} for the polynomial ideal $\mathcal{I}$, which can be constructed by a well-known algorithm called \textit{Buchburger's algorithm}~\cite{Cox15, Cox04, Froeberg97, Becker93}. 
The Gr\"obner basis is not unique, but we can obtain an unique minimal Gr\"obner basis --the \textit{reduced Gr\"obner basis}--  by adding the following restrictions: for each $j=1,2,\cdots,t$, every polynomial $g_j$ is monic and its leading monomial $\text{LM}(g_j)$ is not divisible by  $\text{LM}(g_i)$ for any $i\neq j$. 
Let us denote the reduced Gr\"obner basis for $\mathcal{I}$ by $\mathscr{G}(\mathcal{I})$. 
In the following, when we simply write Gr\"obner basis, we will always refer to a reduced Gr\"obner basis.
The Gr\"obner basis is useful since the corresponding system of polynomial equations:
\begin{align*}
g_1(\theta_1,\cdots,\theta_M)=\cdots=
g_t(\theta_1,\cdots,\theta_M)=0
\end{align*}
has the same zeros as the original system of polynomial equations in Eq.~(\ref{eq:systemeq1}), and usually has a simpler form.

The solvability of the system of polynomial equations over $\mathbb{C}$ depends on the shape of the Gr\"obner basis as follows.
\begin{enumerate}
\item \textbf{No solution}~\cite{Cox15}:
When Eq.~(\ref{eq:systemeq1}) is not solvable, Hilbert's weak Nullstellensatz forces $\mathscr{G}(\mathcal{I})=\{1\}$.

\item \textbf{Finite set of solutions}~\cite{Cox04, Froeberg97}: 
When Eq. (\ref{eq:systemeq1}) has finite solvability (a finite number of solutions), $\mathcal{I}$ is called zero-dimensional ideal. 
With lexicographic order, $\mathscr{G}(\mathcal{I})$ has the shape:
\begin{align*}
\begin{split}
\mathscr{G}(\mathcal{I})=\{&g_1(\theta_1),\\
&g_{2,1}(\theta_1,\theta_2),\cdots,g_{2,v_{2}}(\theta_1,\theta_2),\\
&\vdots\\
&g_{M,1}(\theta_1,\cdots,\theta_M),\cdots,g_{M,v_M}(\theta_1,\cdots,\theta_M)\}. 
\end{split}
\end{align*}
This allows all the values of the parameters to be similarly obtained recursively. In particular, when $\mathcal{I}$ is a radical zero-dimensional ideal, the Gr\"obner basis has a particular shape (Shape lemma):
\begin{align*}
\mathscr{G}(\mathcal{I})=\{\theta_1^{\alpha} +q_1(\theta_1),\theta_{2}+q_{2}(\theta_1), \cdots, \theta_M+q_M(\theta_1)\},
\end{align*}
where $q_j(\theta_1)$ is an univariate polynomial in $\theta_1$ with the condition that $\alpha>\text{deg}(q_j)$ for $\alpha\in\mathbb{N}$. From Sturm theorem~\cite{Cox04}, we can obtain the number of distinct real zeros of $\theta_1^{\alpha}+q_1(\theta_1)=0$ and hence the number of
real solutions of the original system of polynomial equations.

\item{\textbf{Only one solution}~\cite{Froeberg97}:
When Eq.~(\ref{eq:systemeq1}) has only one solution, the radical of the zero-dimensional ideal is the maximal ideal, which has the form of $\langle \theta_1-c_1,\cdots, \theta_M-c_M\rangle$. Therefore, the Gr\"obner basis for $\sqrt{\mathcal{I}}$ has the form 
\begin{align*}
\mathscr{G}(\sqrt{\mathcal{I}})=\{\theta_1-c_1,\cdots,\theta_M-c_M\}.
\end{align*}
}
\end{enumerate}
Buchberger's algorithm for computing the reduced Gr\"obner basis has already been implemented in many commercial softwares.

\section{Identifiability test }
\label{sec:test}
We can now use the Gr\"obner basis formalism to introduce a working definition of Hamiltonian identifiability via the ERA technique. 
The concept of identifiability has been studied in several different contexts~\cite{Burgarth12, Burgarth14, Guta13}. Gu\c{t}\v{a} and Yamamoto~\cite{Guta13} employed a transfer function approach to systematically study system identifiability of the linear quantum systems with continuous variables. Their result applies to continuous-variable quantum systems in \textit{infinite-dimensional} Hilbert space, such as a quantum optomechanical system~\cite{James08, Yamamoto14} or atomic ensembles confined in an optical cavity~\cite{Li09}. 
However, here we are interested in interacting many-body spin-1/2 systems that can be described by  discrete, \textit{finite-dimensional} Hilbert spaces. Since the algebraic structure  of the spin operators is different, we have to reformulate the conditions for identifiability of many-body spin-1/2 Hamiltonians. 

In particular, we focus on  Hamiltonian identifiability for many-body spin-1/2 systems, to provide a procedure to test  identifiability. In addition, we restrict ourselves to identifying only the parameter magnitude,  $|\theta_j|$ . 

Let us first introduce our definition  of Hamiltonian identifiability: 
\begin{defi}
An Hamiltonian is identifiable when the system of polynomial equations derived from the transfer function equation $T(s)= T_{\text{est}}(s)$ provided by ERA has a finite set of solutions such that all  $\theta_j^2$ take only one positive real value.   
\end{defi}

Let $F$ be a polynomial set $F=\{f_1,\cdots,\ f_p\}\subseteq\mathbb{C}[\theta_1,\cdots,\theta_M]$. Based on this definition, the {algorithm} to test identifiability is as follows:

\begin{enumerate}[leftmargin=3pt,itemindent=*, label={\bfseries Step \arabic*:}]
\item We define new variables $z_j$ such that $\{z_j\}=\{\theta_r^2, \theta_l |1\le r,l\le M)\}$, where $\theta_r$'s only appear in $F$ as even powers (and  $\theta_l$'s are all the remaining variables in $F$). Then, the polynomial ideal generated by $F$ becomes $\mathcal{I}=\langle f_1,\cdots, f_p\rangle\subseteq\mathbb{C}[z_1,\cdots, z_M]$. 
\item From the Buchberger's algorithm and the definition of reduced Gr\"obner basis, we obtain $\mathscr{G}(\mathcal{I})=\{g_1,\cdots,g_t\}$.
If $t< M$, the Hamiltonian is non-identifiable. 
\item By elimination of variables, we can obtain $M$ univariate polynomials $h_j(z_j)$, i.e. $h_j\in\mathcal{I}\cap\mathbb{C}[z_j]$. 

Then, we can construct the radical ideal  $\sqrt{\mathcal{I}}=\mathcal{I}+\langle \varphi_1,\cdots,\varphi_M\rangle,
$ where $\varphi_j=h_j/\text{gcd}(h_j, \partial_{z_j}h_j)$
\cite{Cox04, Cox15}, and we can construct a new polynomial set  $F'=\{g_1,\cdots,g_t, \varphi_1,\cdots,\varphi_M \}\subseteq\mathbb{C}[z_1,\cdots, z_M]$.
\item Since $\sqrt{\mathcal{I}}$ is a radical zero-dimensional ideal, the shape lemma can be applied. From  Buchberger's algorithm and the definition of reduced Gr\"obner basis, we obtain:
\begin{align*}
\mathscr{G}(\sqrt{\mathcal{I}})=\{z_1^{\alpha}+q_1(z_1), z_2+q_2(z_1), \cdots, z_M+q_M(z_1)\},
 \end{align*}
 where $\alpha>\text{deg}(q_j)$.

\item  Finally, we employ Sturm theorem to calculate the distinct number of real zeros of the polynomial $z_1^{\alpha}+q_1(z_1)$, so that we can obtain the number of real zeros of each polynomial in $\mathscr{G}(\sqrt{\mathcal{I}})$. If there is only one set of solutions such that all the values of $\theta_j$'s are real, the Hamiltonian is identifiable. Otherwise, the Hamiltonian is non-identifiable. 
\end{enumerate}

\section{Lower bound in number of sampling points}
\label{sec:boundpoint}
In addition to providing an operational definition of Hamiltonian identifiability, ERA together with the Gr\"obner basis technique provides a lower bound for the number of sampling points required to identify all  parameters. The  bound is found from the minimum realization of the system~\cite{Katayama06}. In order to obtain the new realization $[\tilde{\mathbf{A}}_{\text{est}},\mathbf{C}_{\text{est}},\mathbf{x}_{\text{est}}(0)]$, the system is required to be both observable and controllable (see Appendix~\ref{sec:Hankel}). We  thus have
\begin{align*}
\text{rank}(\mathcal{O}_r)=\text{rank}(\mathcal{C}_s)
=\text{rank}(\mathbf{A})=n,
\end{align*}
where $\mathcal{O}_r$ and $\mathcal{C}_s$ are the observability and controllability matrix~\cite{Katayama06}.
Since the Hankel matrix $\mathbf{H}_{rs}(0)$ has the form  $\mathbf{H}_{rs}(0)=\mathcal{O}_r\mathcal{C}_s$, from  Sylvester inequality~\cite{Horn13}, we find that 
\begin{align*}
\text{rank}(\mathbf{H}_{rs}(0))=\text{rank}(\mathbf{A})=n.
\end{align*}
Therefore, the minimum dimension of the Hankel matrix is $n\times n$. Taking into account the need of constructing the shifted Hankel matrix $\mathbf{H}_{rs}(1)$ to obtain $\mathbf{A}_{\text{est}}$, the lower bound in the number of sampling points is:
\begin{align*}
\lambda_{\text{min}}=2\,\text{rank}(\mathbf{A})=2n.
\end{align*}

Since the number of different polynomial equations obtained from Eq.~ (\ref{eq:identity}) is  $\leq\text{rank}(\mathbf{A})-1$, we can also obtain the relation between the lower bound in the number of sampling points and the Gr\"obner basis. Let $\mathcal{N}[\mathscr{G}(\mathcal{\sqrt{I}})]$ be the number of elements of the Gr\"obner basis $\mathscr{G}(\sqrt{\mathcal{I}})$: we can usually write $\mathcal{N}[\mathscr{G}(\sqrt{\mathcal{I}})]\le \text{rank}(\mathbf{A})-1$. Since $\lambda_{\text{min}}=2\,\text{rank}(\mathbf{A})$, we have:
\begin{align*}
\frac{\lambda_{\text{min}}}{2}\ge \mathcal{N}[\mathscr{G}(\sqrt{\mathcal{I}})]+1. 
\end{align*}
From the measurement number, we can further obtain the time required for Hamiltonian identification.
The optimal choice of the time interval $\Delta t$ is given by the Sampling theorem~\cite{Shannon49}. 
Let $\Omega_{\text{max}}/(2\pi)$ be the maximum frequency that would appear in the measured signal. Then, $\Delta t$ has to satisfy $\Delta t\le \pi/\Omega_{\text{max}}$. 
Therefore, the required maximum evolution time with the minimum number of sampling points satisfies: 
\begin{align*}
t_{\text{tot}}\le \frac{(2n-1)\pi}{\Omega_{\text{max}}}.
\end{align*}

In reality, the maximum frequency of the signal depends on the values of the parameters $\theta_m$, which are unknown. Thus, for a time-optimal estimation procedure we would need prior information about the range of values that the parameters can take. For example, we could then assume that all the parameters take the largest value and obtain the smallest time steps $\Delta t$ that still satisfy the sampling theorem.

%\pc{note: the time found here is not consistent with the later discussion of minimum time. Please correct here, and use $\omega$ everywhere instead of $\sigma$ to designate the frequency.}

\section{Examples: Hamiltonian identifiability test}
\label{sec:examples}

\begin{figure}[bht]
\includegraphics[width=0.45\textwidth]{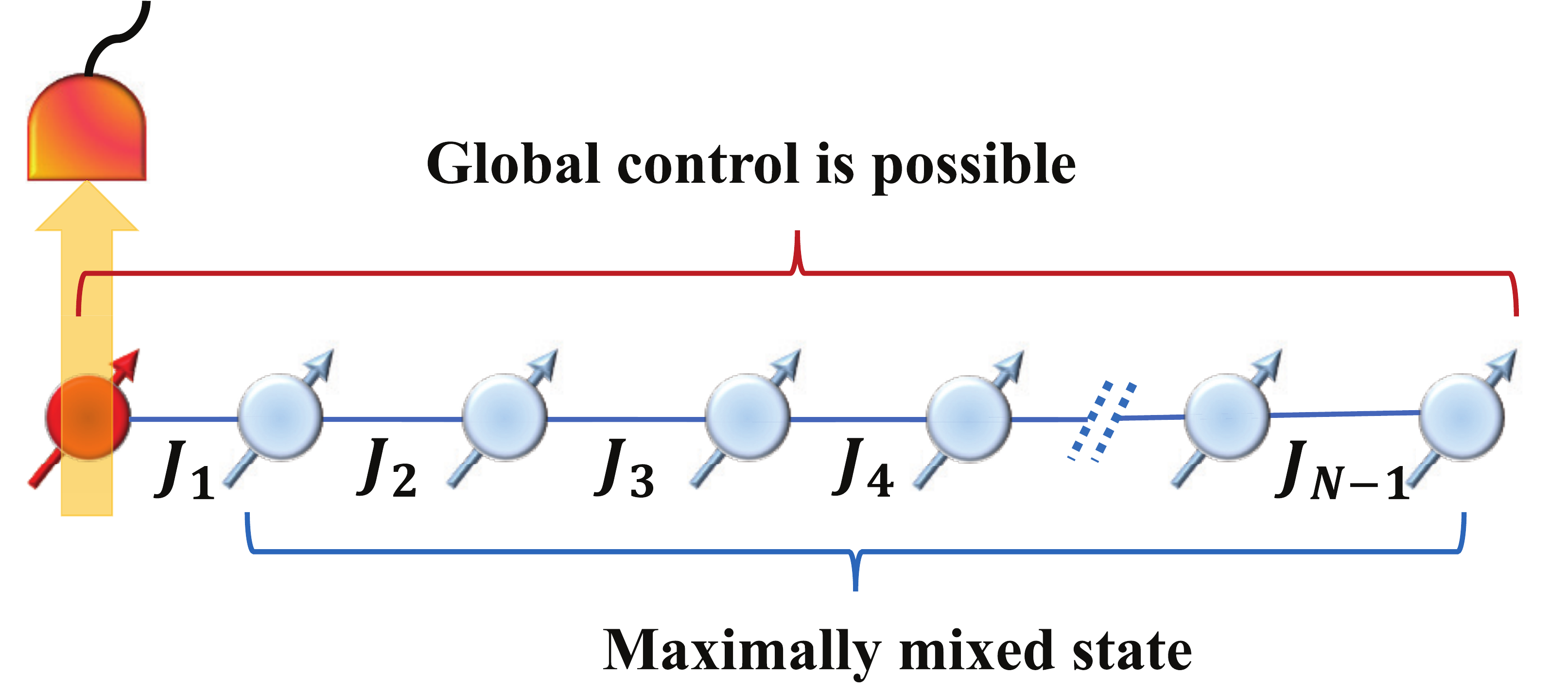}
\caption{
\textbf{Hamiltonian identification model}. A quantum probe is coupled to one end of the spin chain. 
A part from the quantum sensor (red spin), the rest of the spins (blue spins) are initially in the maximally mixed state. We further assume that we only have  selective control on the quantum probe and  global control on the spin chain.}
\label{fig:model}
\end{figure}

%\pc{when using the \slash~label\{\} command, use a label that can be easily recognized and remembered (not sec 1, 2 etc., so that if things are moved around the label still make sense)}
We now presents some exemplary systems and analyze their identifiability, as well as the minimum number of  sampling points and time for  Hamiltonian identification. 
To provide  analytical results, we focus our attention on nearest-neighbor coupling spin chains, which is a useful model for quantum information transport between distant qubits~\cite{Bose03, Franco08, Cappellaro11}.  We consider two different Hamiltonians, the Ising and exchange models, and analyze their Hamiltonian identifiability by assuming that the spin chain is coupled to a single quantum probe. More precisely, we make the following assumptions (see also Fig.~\ref{fig:model}). 

{\itshape\begin{enumerate}
\item{The quantum probe can be initialized, controlled, and read out. The quantum probe is coupled to the chain by one of its  end spin with a coupling that follows the chain Hamiltonian model.}
\item{The chain spins cannot be initialized nor measured. For simplicity, we thus assume that they are initially in the maximally mixed state $\frac{1}{2}\openone$. We only allow collective control on the chain spins.}  
%\item{The initial state of the quantum probe is chosen such that the initial coherent vector $\mathbf{x}(0)$ is non-zero.}
\item{The coupling type and the size of the spin chain are known.}  
\end{enumerate}}
These assumptions are realistic in many practical scenarios for 
spin chain system applications to quantum engineering tasks at room temperature. In addition, they could also approximate some scenarios in quantum metrology, such as recently proposed schemes for protein structure reconstructions via the interaction of their nuclear spins with a quantum probe~\cite{Ajoy15}.

For concreteness, we consider a spin-$1/2$  chain  comprising $N$ spins (including the quantum probe) with nearest-neighbor interactions and possibly an interaction to an external field.
The parameters $\theta_m$ in Eq.~(\ref{eq:Hamiltonian1}) are thus given by 
the coupling strengths between the $k$-th and $(k+1)$-th spins,  denoted by $J_k/2$,  and the Zeeman energy $\omega_k/2$ of the $k$-th spin due to  external fields. 

\subsection{Ising model without transverse field}
\label{sec:ising}
As a preliminary example of the methods, we consider  the Ising model without transverse field: 
\begin{equation}
H=\sum_{k=1}^{N-1}\frac{J_k}{2}S_{k}^{\alpha}S_{k+1}^{\alpha}.
\label{eq:ising}
\end{equation} 
For concreteness, we can select $S^{\alpha}=X$  and $G_0=\{Z_1\}$ without loss of generality. The accessible set is easily obtained from the commutators and it saturates very quickly: 
\begin{align*}
G=\{Z_1, Y_1 X_2\}.
\end{align*} 
Then, only the spin directly interacting with the quantum probe becomes correlated with it during the evolution and its parameter can be identified. As a consequence, $\text{rank}(\mathbf{A})=2$ and full Hamiltonian identification is  only possible for $N=2$.  
Physically, this can be understood by a lack of information propagation in the Ising spin chain, which prevents the quantum probe at its end to gain information about the rest of the system. Indeed, the group velocity for information propagation in the Ising chain is zero. 

Let the initial coherent vector be $\mathbf{x}(0)=(1,0)^{T}$ and the output matrix $\mathbf{C}=\begin{pmatrix}1&0\end{pmatrix}$. Then, the transfer function is:
\begin{align*}
T(s)=\frac{s}{s^{2}+J_1^{2}},
\end{align*}
where we can identify $z_1=J_1^2$. Through ERA, we can obtain a new transfer function from the experimental data, which can be written in the most generality as
\begin{align*}
T_{\text{est}}(s)=\frac{s+b_0}{s^2+b_1^2s+a_1^2}.
\end{align*}
Here we fixed  the transfer function order to 2, as expected from the Ising model evolution. However, the form of $T_{\text{est}}(s)$ might  differ from the ideal $T(s)$: in particular we might have additional terms, with coefficients  $b_j$ arising from experimental errors or numerical approximations. Since $a_1^{2}$ reflects the contribution of $J_1^{2}$, we still expect $b_j^{2}\ll a_1^{2}$, and $b_j$'s are negligible. 
%\pc{and so they are neglected?}
Therefore, the Gr\"obner basis is simply given by: $\mathscr{G}=\{z_1-a_1^2\}=\{J_1^2-a_1^2\}$ and the $N=2$ Ising chain can be identified with $\lambda_{\text{min}}=4$ sampling points.

We note that an alternative way of estimating $J_1$ is to measure the quantum probe (in particular $Z_1$) at known times. However, due to the periodicity of the signal, $J_1$ cannot be identified uniquely, even if measuring more than one time point.

We can thus generally state the following result.
\begin{result}
The Ising model without the transversefield is only identifiable via the measurement of a sinlge probe spin for $N=2$, and the lower bound in the number of sampling points is given by: $\lambda_{\text{min}}=2\,\text{rank}(\mathbf{A})=4$.
\end{result}

\subsection{Ising model with transverse field}
\label{sec:isingtr}
Adding a transverse field to the Ising model drastically changes the system dynamics and consequently its identifiability.

The Hamiltonian is now 
\begin{align*}
H=\sum_{k=1}^{N}\frac{\omega_k}{2}S^{\gamma}_k+\sum_{k=1}^{N-1}\frac{J_k}{2}S_{k}^{\alpha}S_{k+1}^{\alpha},
\end{align*}
where for concreteness we will set $S^\alpha=X$ and $S^\gamma=Z$. 
There are several possible observable sets $G_0$ to choose from, as none of the operators $S^\xi_1$ ($\xi=\{\alpha,\beta,\gamma\}$) commute with the Hamiltonian.
For the case considered, setting either $G_0=\{X_1\}$ or $\{Y_1\}$ is the best choice, as $G_0=\{Z_1\}$ would result in a larger-size $\tilde{\mathbf{A}}$. 
Either $G_0=\{X_1\}$ or $\{Y_1\}$  yields the accessible set: 
\begin{align*}
\begin{split}
G=\{X_1, Y_1, &Z_1X_2, Z_1Y_2,\cdots,\\
&Z_1\cdots Z_{N-1}X_N, Z_1\cdots Z_{N-1}Y_N\},
\end{split}
\end{align*}
so that $\text{dim}(G)=2N$.
All the chain spins are thus correlated with the quantum probe and we can  hope to identify all the parameters. 
The system matrix $\tilde{\mathbf{A}}$ is 
a $2N\times 2N$ skew-symmetric matrix with the only non-zero elements $\tilde{\mathbf{A}}_{2k,2k-1}=\omega_k$ and $\tilde{\mathbf{A}}_{2k+1,2k}=J_k$. Since $\text{dim}(G)=\text{rank}(\mathbf{A})$, we have: $\text{rank}(\mathbf{A})=2N$. 
Choosing the initial state of the quantum probe to be the eigenstate of $X_1$, we have:
\begin{align*}
\mathbf{x}(0)=(1,0\cdots,0)^{T}\in\mathbb{R}^{2N}
\end{align*}
If we measure $X_1$, the output matrix is $\mathbf{C}=\begin{pmatrix}1&0&\cdots&0\end{pmatrix}\in\mathbb{R}^{2N}$. From ERA and Eq.~(\ref{eq:identity}), we  arrive at the following shape of the Gr\"obner basis: 
\begin{align*}
\mathscr{G}=\{\omega_{1}^{2}-a_{1}^{2}\cdots, \omega_{N}^{2}-a_{N}^{2},J_1^{2}-b_1^{2}, \cdots, J_{N-1}^{2}-b_{N-1}^{2}\}
\end{align*}
because in this case  $\mathcal{I}$ generated from Eq.~(\ref{eq:identity}) is a maximal ideal of the form 
$\langle z_1-a_1^2,\cdots, z_N-a_N^2, z_{N+1}-b_1^2, \cdots, z_{2N-1}-b_{N-1}^{2}\rangle\ (a_l,b_k\in\mathbb{R})$,
where $z_l=\omega_{l}^{2} \ (l=1,\cdots,N)$ and $z_{N+k}=J_{k}^2 \ (k=1,\cdots,N-1)$.  
Since there is only one positive real solution for the magnitudes of all the parameters, the Hamiltonian is fully identifiable.

If we measure $Y_1$ with the initial state of the quantum probe being the eigenstate of $X_1$, the Gr\"obner basis is instead:
\begin{align*}
\begin{split}
\mathscr{G}=\{\omega_{1}-a_{1},\ &\omega_{2}^{2}-a_{2}^{2},\cdots, \omega_{N}^{2}-a_{N}^{2},\\
&J_1^{2}-b_1^{2}, \cdots, J_{N-1}^{2}-b_{N-1}^{2}\},
\end{split}
\end{align*}
showing that we can find the sign of $\omega_1$, in addition to identifying the magnitude of all other parameters. 

Physically, this result shows that identifiability is connected to information propagation along the whole chain. Indeed, since we assumed that we can extract information from the system only through the probe spin at one end of the chain, propagation of information through the whole chain is necessary to 
reveal the system's properties. Adding a transverse field to the Ising model enables this information propagation.

We can thus generally state the following result.
\begin{result}
	The Hamiltonian of the nearest-neighbor Ising model with transverse field  is identifiable via  measurement of a single quantum probe. The minimum number of sampling points for $N$ spins is  $\lambda_{\text{min}}=2\text{rank}(\mathbf{A})=4N$. 
\end{result}

\subsection{Exchange model without transverse field}
\label{sec:exchange}
The exchange (XY) model is another example where information propagation allows Hamiltonian identification via single-probe measurement.

The Hamiltonian can be written as: 
%\pc{it's "written as" not "written by"}
\begin{align}
H=\sum_{k=1}^{N-1}\frac{J_{k}}{2}(S^{\alpha}_k S^{\alpha}_{k+1}+S^{\beta}_k S^{\beta}_{k+1}),
\label{eq:exchange}
\end{align}
where for concreteness we will set $S^\alpha=X$ and $S^\beta=Y$. For this case, $G_{0}=\{X_{1}\}$ or $\{Y_1\}$ is the best choice because the corresponding accessible set has the smallest size. Choosing, e.g.,  $G_0=\{X_1\}$,  we  obtain the following accessible set 
\begin{align*}
\begin{split}
G=\{X_1,&Z_1 Y_2, Z_1 Z_2 X_3, Z_1 Z_2 Z_3 Y_4, \cdots,\\
&Z_1\cdots Z_{2m-2}X_{2m-1}, Z_1\cdots Z_{2m-1}Y_{2m}\},
\end{split}
\end{align*}
for an even-number of spins in the chain, $N=2m~(\forall m\in\mathbb{N})$, and 
\begin{align*}
\begin{split}
G=\{X_1,&Z_1 Y_2, Z_1 Z_2 X_3, Z_1 Z_2 Z_3 Y_4, \cdots,\\
&Z_1\cdots Z_{2m-3}Y_{2m-2}, Z_1\cdots Z_{2m-2}X_{2m-1}\},
\end{split}
\end{align*}
for an odd number, $N=2m-1~(\forall m\in\mathbb{N})$.

The accessible set has the smallest possible dimension,  $\text{dim}(G)=N$. (If we had chosen $G_0=\{Z_1\}$, the accessible set dimension would have been $\text{dim}(G)=N^2$. Therefore, in the following discussion, we consider $G_0=\{X_1\}$.) %\pc{is all the following for G0=X? I was confused...}
As all the spins are correlated with the quantum probe, we can expect the Hamiltonian to be fully identifiable. 
 The system matrix $\tilde{\mathbf{A}}$ becomes an $N\times N$ skew-symmetric matrix with the only non-zero elements $(\tilde{\mathbf{A}})_{k,k+1}=(-1)^{k}J_{k}$, which has the same form as the system matrix of the Ising model with the transverse field. Since $\text{dim}(G)=\text{rank}(\mathbf{A})$, we have $\text{rank}(\mathbf{A})=N$. Choosing the initial state of the quantum probe to be the eigenstate of $X_1$, we have:
\begin{align*}
\mathbf{x}(0)=(1,0,\cdots,0)^{T}\in\mathbb{R}^{N}.
\end{align*}
Since we can only measure the quantum probe, the output matrix is $\mathbf{C}=\begin{pmatrix}1&0&\cdots&0\end{pmatrix}\in\mathbb{R}^{N}$. From ERA and Eq.~(\ref{eq:identity}), we  arrive at the following shape of the Gr\"obner basis:
\begin{align*}
\mathscr{G}=\{J_1^{2}-a_1^{2},\cdots, J_{N-1}^{2}-a_{N-1}^{2}\},
\end{align*} 
where $a_k\in\mathbb{R}$. Therefore, the Hamiltonian is fully identifiable since we have only one positive real solution for the magnitudes of all the parameters. 

We can thus generally state the following result.
\begin{result}
The  Hamiltonian  of the nearest-neighbor exchange model without  transverse field  is identifiable via  measurement of a single quantum probe. The minimum number of sampling points for $N$ spins is  $\lambda_{\text{min}}=2\text{rank}(\mathbf{A})=2N$.
\end{result}

\subsection{Exchange model with transverse field}
\label{sec:exchangetr}
Adding a transverse field to the exchange Hamiltonian complicates the situation, as there might be more than one solution to the identification problem. However, this can be resolved by performing the measurements on two different observables.  

The Hamiltonian is now given by:
\begin{align*}
H=\sum_{k=1}^{N}\frac{\omega_k}{2}S^{\gamma}_k+\sum_{k=1}^{N-1}\frac{J_{k}}{2}(S^{\alpha}_k S^{\alpha}_{k+1}+S^{\beta}_k S^{\beta}_{k+1}),
\end{align*}
where for concreteness we choose  $S^{\alpha}=X$,  $S^{\beta}=Y$, and $S^{\gamma}=Z$. 
Again, the best choice  is either $G_0=\{X_1\}$ or $G_0=\{Y_1\}$ that both yield the following accessible set:
\begin{align*}
\begin{split}
G=\{X_1, Y_1, Z_1 X_2, &Z_1 Y_2, Z_1Z_2X_3, Z_1Z_2Y_3,\cdots,\\ &Z_1\cdots Z_{N-1}X_N, Z_1\cdots Z_{N-1}Y_N\},
\end{split}
\end{align*}
with $\text{dim}(G)=2N$.  The system matrix is a $2N\times 2N$ skew-symmetric matrix with the only non-zero elements $(\tilde{\mathbf{A}})_{2k-1,2k}=\omega_{k}$ and $(\tilde{\mathbf{A}})_{2k+2,2k-1}=(\tilde{\mathbf{A}})_{2k,2k+1}=J_k$. 
Choosing the initial state of the quantum probe to be an eigenstate of $X_1$, we have:
\begin{align*}
\mathbf{x}(0)=(1,0\cdots,0)^{T}\in\mathbb{R}^{2N}.
\end{align*}
If we measure $X_1$, then we have $\mathbf{C}=\begin{pmatrix}1&0&\cdots&0\end{pmatrix}\in\mathbb{R}^{2N}$. From ERA and Eq. (\ref{eq:identity}), we can construct a zero-dimensional ideal radical $\sqrt{\mathcal{I}}$. From the shape lemma, we  obtain the following Gr\"obner basis:  
\begin{align*}
\begin{split}
\mathscr{G}(\sqrt{\mathcal{I}})=\{&z_1^{\alpha} +q_1(z_{1}),\\ 
&z_2+q_2(z_1),\\ 
&\vdots\\
&z_{2N-1}+q_{2N-1}(z_{1})\},
\end{split}
\end{align*}
where  $z_l=\omega_{l}^{2} \ (l=1,\cdots,N)$ and $z_{N+k}=J_{k}^2 \ (k=1,\cdots,N-1)$. Note that $\alpha>\text{deg}(q_j)$ and $\alpha\ge 2$. Here, $q_j(z_1)$ %\pc{it was g??} 
is the univariate polynomial in $z_1$. In general $z_{1}$ could have multiple values, so that we could have multiple sets of real solutions of the system of polynomial equations. Therefore, in general, this model is not identifiable. If we measure $Y_1$ with the initial state of the quantum probe being the eigenstate of $X_1$, we also have the same situations. Therefore, the exchange model with transverse field is generally not identifiable if we only measure one observable. %\pc{what if you measure both X1 and Y1? what if you also measure Z1?}

This issue  can be resolved by measuring two different basis operators. Suppose that we measure $X_{1}$ and $Y_{1}$ with initial coherent vector $\mathbf{x}(0)=(1,0,\cdots,0)^{T}\in\mathbb{R}^{2N}$. In this case, we collect the measurement data for two observables, so that the sampling matrix $\mathbf{C}$ becomes: $\mathbf{C}=\begin{pmatrix}1&1&\cdots&0\end{pmatrix}\in\mathbb{R}^{2N}$. Then the transfer function can be written as the sum of the one for $X_1$ and the one for $Y_1$:
\begin{align*}
T(s)=T^{(X_{1})}(s)+T^{(Y_{1})}(s),
\end{align*}
where $T^{(X_{1})}(s)$ and $T^{(Y_{1})}(s)$ have  order  $2N$. Therefore, the order of the transfer function $T(s)$ is still $2N$. In order to obtain the new realization, we perform the singular value decomposition of two Hankel matrices,  corresponding to $X_1$ and $Y_1$, respectively. Thus, we can obtain the following new transfer function:
\begin{align*}
T_{\text{est}}(s)=T_{\text{est}}^{(X_{1})}(s)+T_{\text{est}}^{(Y_{1})}(s).
\end{align*}
From the identity $T(s)=T_{\text{est}}(s)$, the polynomial ideal turns out to be a maximal ideal, which has the form of:
\begin{align*}
\begin{split}
\mathcal{I}=\langle z_{1}-a_{1},&\cdots, z_{N}-a_{N},\\
&z_{N+1}-b_{1}^{2},\cdots, z_{2N-1}-b_{N-1}^{2} \rangle,
\end{split}
\end{align*}
where $z_{k}=\omega_{k}$ and $z_{N+k}=J_{k}^{2}$ and $a_{k},b_{k}\in\mathbb{R}$. Therefore, the Gr\"obner basis becomes:
\begin{align*}
\begin{split}
\mathscr{G}(\mathcal{I})=\{\omega_1-a_1,&\cdots,\omega_{N}-a_{N},\\
& J_{1}^{2}-b_{1}^{2},\cdots, J_{N-1}^{2}-b_{N-1}^{2}\}.
\end{split}
\end{align*}
The Hamiltonian is now fully identifiable since we have only one positive real solution for the magnitudes of all the parameters, and in addition we can find the sign of $\omega_k$. 

Note that in this case, since we need two measurements, we need $2\times2\text{rank}(\mathbf{A})=8N$ sampling points in total. This result can be understood as  follows.  The information provided by the time evolution of only one observable is not sometimes enough to extract the exact values of the parameters, but we can obtain a set of possible solutions. However, additional information provided by different observables can allow us to exclude some solutions. For the exchange model with transverse field, we can restrict the set of  solutions to  only one solution by adding the information provided by $Y_1$ to the information provided by $X_{1}$. Hence, we can generally state the following result.
\begin{result}
The Hamiltonian of the nearest-neighbor exchange model with  transverse field  is generally non-identifiable via the measurement on a single quantum probe if we only measure one observable. If we observe two observables, the Hamiltonian can be fully identified and furthermore we can determine the sign of the Zeeman splitting. In this case, the minimum number of sampling points for $N$ spins is  $\lambda_{\text{min}}=4\text{rank}(\mathbf{A})=8N$.
\end{result}
%\pc{don't you have a better way to express the condition when the system has just one solution? (as expressed just above...)}

\subsection{Time required for identification of spin chains}
\label{timebound}

By analyzing ERA procedure, we obtained bounds on the minimum number of sampling points required for Hamiltonian identification. In turns, this also leads to requirements on the minimum evolution time as well as the total time required for Hamiltonian identification. 

Indeed, if some \textit{a priori} information about the system is known, we can choose the maximum time step required by the sampling theorem, $\Delta t=\pi/\Omega_{\text{max}}$, where $\Omega_{\text{max}}$ is the maximum eigenvalue of the Hamiltonian. With the minimum number of  sampling points, the longest evolution time is at lest $t_{\text{tot}}=(\lambda_{min}-1)\Delta t$: This time should be compared to the system coherence time. 
In addition, the overall Hamiltonian identification requires a time 
\begin{equation}
t_{\text{id}}=\Big(\frac{\lambda_{min}-1}{2}\Delta t+t_{\textrm{dead}}\Big)\lambda_{min},
\label{eq:measurementtime}
\end{equation}
where $t_{\textrm{dead}}$ is the dead time associated with system initialization and readout.

We can further check that these time requirements are consistent with our intuitive physical picture that connects Hamiltonian identification to the propagation of  information through the whole spin chain. Consider for example the exchange Hamiltonian, Eq.~(\ref{eq:exchange}), with all equal couplings. 
We can assume that the spins are equally spaced, with $a$  the lattice constant, and $L=(N-1)a$  the length of the chain. 
The eigenvalues of the Hamiltonian in the first excitation manifold are then
\begin{align*}
\Omega_{n}=2J\cos\Big(\frac{n\pi}{N+1}\Big)=2J\cos(k_na), \quad k_n=n\pi\frac{N-1}{L(N+1)},
\end{align*}
with $n=1,2,\dots,N$. Then, the maximum angular frequency is 
\begin{align*}
{\Omega_{\text{max}}}=2{J}\cos\Big(\frac{\pi}{N+1}\Big).
\end{align*}
Since we need at least $2N$ sampling points, the longest evolution time is $t_{\text{tot}}=(2N-1)\pi/\Omega_{\text{max}}$. For large $N\gg 1$, we can simplify this to
\begin{align*}
t_{\text{tot}}\simeq\pi\frac{N}{J}.
\end{align*}

In order for the quantum probe to extract information on the whole spin chain, information needs to propagate to the other end and back. 
We can compute the group velocity for the propagation of the initial excitation on the first (probe) spin from the Hamiltonian eigenvalues $\Omega(k)$~\cite{Ramanathan11}, 
\begin{align*}
v_{g}=\Big|\frac{d\Omega(k)}{dk}\Big|_{\text{max}}=2Ja=\frac{2JL}{N-1}.
\end{align*}
{Then}, the time required for the information to come back to the probe spin is approximately given by
\begin{align*}
\tau\simeq\frac{2L}{v_g}=\frac{N-1}{J}
\end{align*}
Thus, for large $N\gg 1$, we have
$\tau\simeq \frac{N}{J}$, in agreement with the result obtained from the mathematical requirements for system identification.

\section{Identifiability with external control}
\label{sec:control}
Until now we have analyzed identifiability under the assumption that we can initialize, measure, and control only the probe qubit. We found that some Hamiltonians cannot be identified, since they do not generate enough correlations among the target spins, or equivalently they do not transport information about the probe spin excitation through the whole chain. If we relax these assumptions and allow for a minimum level of control on the target spins, the picture changes. For example, if the target spins can be controlled via collective rotations, it is possible to turn a nonidentifiable Hamiltonian into an identifiable one.

\begin{figure}[htb]
\includegraphics[width=0.45\textwidth]{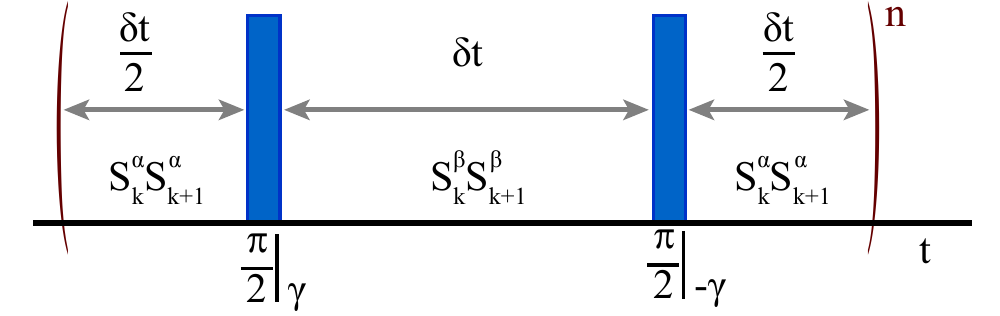}
\caption{\textbf{Identifiability with external control}. By applying a periodic control pulse sequence $n$ times in the limit of a very small $J_k\delta t\ll1$, we can transfer $H_{\text{is}}=\sum_{k=1}^{N-1}\frac{J_k}{2}S_k^{\alpha}S_{k+1}^{\alpha}$ to $H_{\text{ex}}=\sum_{k=1}^{N-1}\frac{J_k}{2}(S_k^{\alpha}S_{k+1}^{\alpha}+S_k^{\beta}S_{k+1}^{\beta})$ so that we can use $2N$ sampling points to identify the parameters $J_k$.}
\label{fig:sequence}
\end{figure}

Consider, for example, the Ising Hamiltonian, Eq.~(\ref{eq:ising}), which we showed in Sec.~\ref{sec:ising} to be nonidentifiable. Using a simple control sequence (see Fig.~\ref{fig:sequence}), we can generate an effective Hamiltonian~\cite{Haeberlen76} that can now be identified, since in the limit of small inter-pulse delays it has the same form as the exchange Hamiltonian, Eq.~(\ref{eq:exchange}). Similarly,  we could use a simple spin-echo procedure to refocus the transverse field and identify the coupling Hamiltonian in Sec.~\ref{sec:exchangetr}, without the need to measure two observables.

More precisely, periodic pulse sequences such as in Fig.~\ref{fig:sequence} make the system evolve as if under an effective time-independent Hamiltonian averaged over the cycle time. The effective Hamiltonian can be approximated   by a first order Magnus expansion~\cite{Magnus54} (average Hamiltonian~\cite{Haeberlen76}). In this limit, to analyze the identifiability, it is sufficient to consider the average Hamiltonian. 
The exact effective Hamiltonian will be identifiable as long as we can identify its approximation; however, its expression might be too complex and analytical results only available in the limit of small enough time interval $\delta t\ll 1$ between the pulses where the approximation holds.
%\pc{not sure why you changed this paragraph. The original version was ok, but the modified not so much. I changed it even further}

{\section{Robustness of ERA Hamiltonian identification}
\label{sec:robustness}
While previous works have already analyzed the robustness of the ERA procedure to experimental errors~\cite{Hou13}, here we want to evaluate the accuracy of the identification algorithm when it is implemented using only the minimum number of measurement points found above. 

To compare with previous results, we consider the {Ising model (with transverse field) for a chain of $N=3$ spins} and {the exchange model (without field) for a chain of $N=6$ spins.} 
We consider the average error in {500} random Hamiltonian realizations and implement the ERA method,  with the minimum number of measurement points ($\lambda_{\text{min}}=4N=12$  for the Ising model and  $\lambda_{\text{min}}=2N=12$ points for the exchange Hamiltonian). We find that the relative error averaged over all the realization is still small ($10^{-10}-10^{-2}~[\%] $) and comparable to previous results, where many more points were measured.

Since the addition of experimental noise could change this result, we study the algorithm robustness in the presence of  noise, 
as a function of  the resources employed during the overall measurement process. 
{To that end, we statistically compare the estimation robustness achieved by using Hankel matrices with different size but keeping fixed the experimental resources. }
 \begin{figure}[t]\centering
\includegraphics[width=0.45\textwidth]{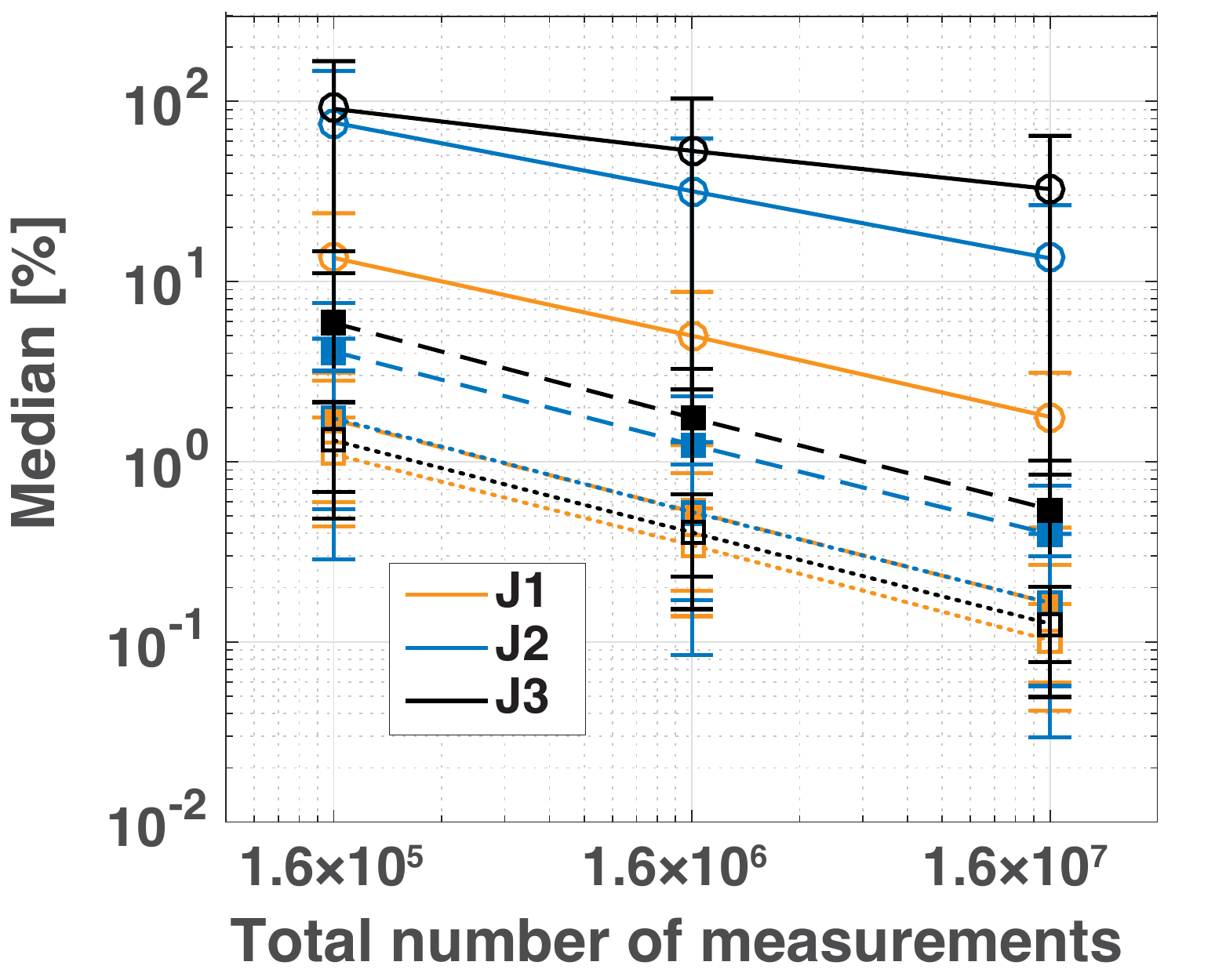}
\caption{\textbf{Estimation error with fixed time step $\Delta t$}. Median of the estimation error $\{\langle\epsilon (J_i)\rangle\}$ over $500$ random Hamiltonian realizations  as a function of the total number of measurements. For each Hamiltonian, we repeated the ERA estimation {$100$} times, to evaluate the {average} error $\langle\epsilon (J_i)\rangle$. Solid lines with circles: $4\times4$ Hankel matrix; dashed lines with solid square: $8\times8$; dotted lines with squares: $40\times40$. {The error bars are the absolute median deviation.} }
\label{fig:FixdtAveError}
\end{figure}

We assume that each sampling point is measured $\mathcal M$ times, yielding a random outcome with a Gaussian distribution $\mathcal N(y(k),\sigma/\sqrt{\mathcal M})$, that is, we assume that the mean is centered around the ``true outcome" value $y(k)$ at each time $k\Delta t$ and for simplicity consider a gaussian noise (with $\sigma=1$). 
By acquiring $2j$ sampling points (with $2j\mathcal M$ total measurements) we can construct  the (noisy) $j\times j$ Hankel matrices $\tilde{\mathbf{H}}_{j}(0)$ and $\tilde{\mathbf{H}}_{j}(1)$. Using the ERA algorithm we can extract a set of parameters $\{\theta_m+\delta \theta_m\}_{m=1}^{M}$ that differ from the true parameters $\{\theta_m\}_{m=1}^{M}$. Since we are interested in the magnitude of the parameters, the estimation error can be written as: 
\begin{align*}
\epsilon(\theta_m)=\Big|\frac{|\theta_m +\delta\theta_m|-|\theta_m|}{|\theta_m|}\Big|\times 100~[\%].
\end{align*}
In the simulations we repeat $r$ times this procedure in order to obtain the {mean} estimation error, $\langle \epsilon(\theta_m)\rangle$ and we further {take the median} over many realizations of the input model parameters.

In our simulation, we compare the estimation errors for Hankel matrices of different sizes $j$, keeping however fixed the total number of measurements, $2j\mathcal M$. The smallest matrix has dimension $n\times n$, where $n$ is the model order. Larger matrices, of dimension $Ln\times Ln$, will thus have an increased error rate by a factor $\sqrt{L}$. 
%with the smallest size and larger size with same total number of measurements for the parameter estimation. We take two size of Hankel matrices: $n\times n$ and $Ln\times Ln$, where $n$ is the model order and $L>1$ is the positive integer larger than $1$. 
Since the presence of the noise forces $\tilde{\mathbf{H}}_{Ln}(0)$ to be full-rank,  we employ low-rank approximation via singular value decomposition~\cite{Markovsky12} to generate an approximated $Ln\times Ln$ Hankel matrix with rank $n$. 

We further consider two scenarios: either the time step $\Delta t$ is fixed (thus larger matrices require longer total times) or the total evolution time $T$ is fixed (reflecting, e.g., constraints imposed by decoherence or experimental drifts). In the first case, $\Delta t$ is chosen by assuming all the parameters take the possible maximal values so that the sampling theorem still holds. 
{In the second case, we fix the total time evolution time to $T=(2n-1) \Delta t$ as required for the smallest Hankel matrix, and we use smaller time steps in the other cases}

As an example, we focus on the $N=4$ exchange model without transverse field, which is shown in Eq.~(\ref{eq:exchange}). The model order is given by $n=4$. Since we assume that the maximal possible value taken by coupling strengths is $100$, we have $dt=\frac{\pi}{25 \sqrt{5}}$.

In Fig.~\ref{fig:FixdtAveError} we plot the estimation errors $\{\langle\epsilon (J_i)\rangle\}$ as a function of the total number of measurement for different Hankel matrix dimensions. 
%We also plot the relation between the standard deviation and the number of measurement repetitions.  

 \begin{figure}[t]\centering
\includegraphics[width=0.45\textwidth]{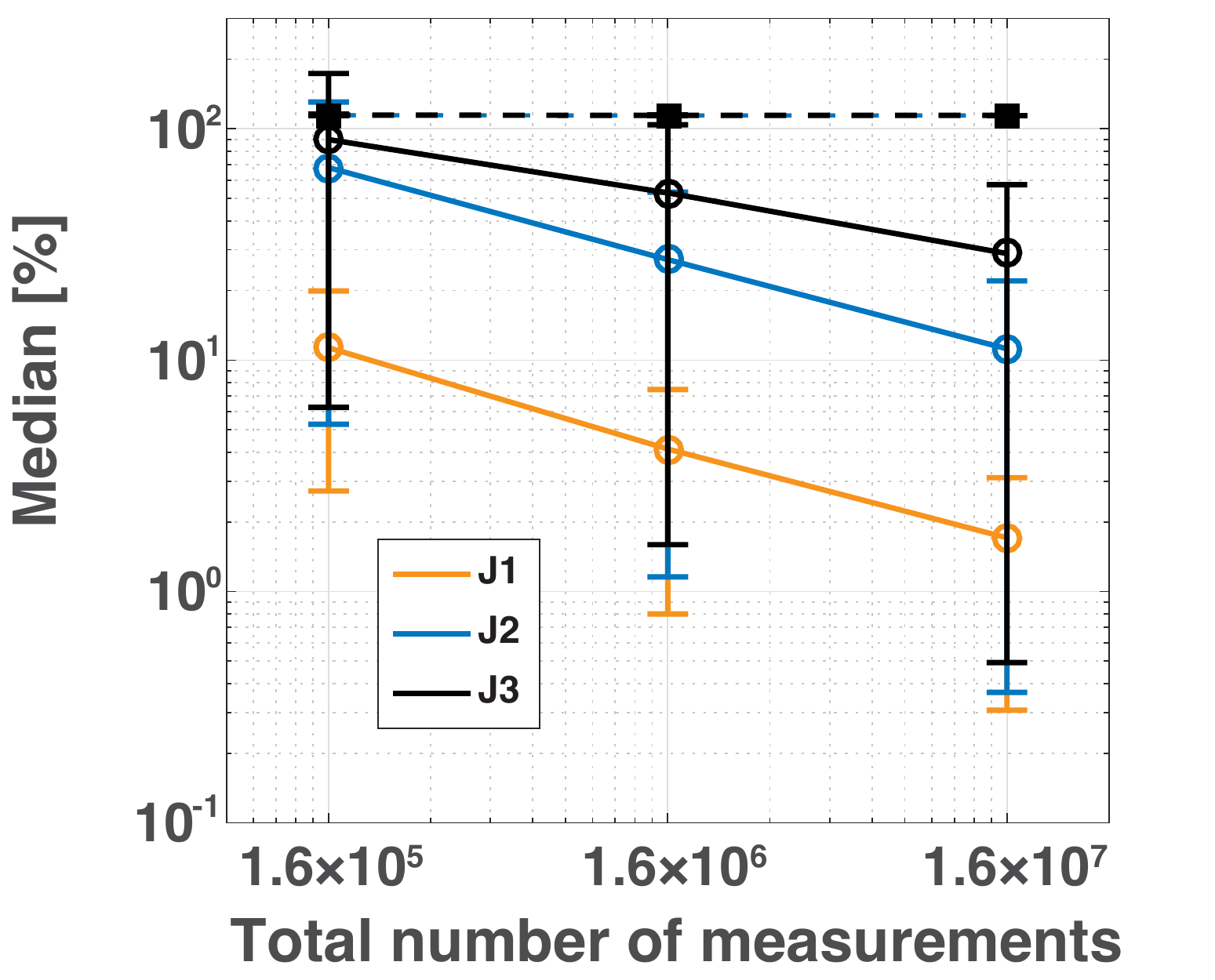}
\caption{\textbf{Estimation error with fixed total time $T$}: {Median of the estimation error $\{\langle\epsilon (J_i)\rangle\}$ over $500$ random realizations of the Hamiltonian as a function of the total number of measurements. For each Hamiltonian, we repeated the ERA estimation $100$ times, to evaluate the averaged error. Solid lines with circles: $4\times4$ Hankel matrix; dashed lines with solid square: $8\times8$. The error bars are the absolute median deviation.} }
\label{fig:FixT4vs8}
\end{figure}

%First, we show the case that $dt$ is fixed for both $4\times 4$ Hankel matrix and $40\times 40$ Hankel matrix.  Then, we obtain Fig.~\ref{fig:FixdtAveError} and Fig.~\ref{fig:FixdtStdError}. 
{We note that the smallest Hankel matrix leads to larger errors, but already slightly larger matrices,  possibly thanks to the low-rank approximation, give more accurate estimation. 
Indeed, thanks to the  low-rank approximation, we generate a four-rank approximation of $\tilde{\mathbf{H}}_{4L}$ by neglecting {the  smallest singular values}, which corresponds to an effective strategy for noise reduction. A second reason for the larger error is related to the shorter total time for the smallest Hankel matrix realization, that might in some cases not allow one to fully capture the smallest frequencies in the signal. While this is not typically an issue in the ideal case, in the presence of experimental noise this leads to higher estimation errors.}
%\begin{figure}[htp!]
%\includegraphics[width=0.54\textwidth]{Fix_dt_AveExchange.eps}
%\caption{\textbf{Average estimation error with fixed time step $dt$}: We plot the relation between the total number of measurements with average estimation error $E(J_k)$ over $100$ random realizations of the Hamiltonian for each size of Hankel matrix. Horizontal axis shows the total number of measurements, where the total number of measurements is given by $8\times\{10^6, 10^7, 10^8, 10^9, 10^{10}\}$. The vertical axis shows the average estimation error $E(J_k)$. }
%\label{fig:FixdtAveError}
%\end{figure}
%\begin{figure}[htp!]
%\includegraphics[width=0.54\textwidth]{Fix_dt_StdExchange.eps}
%\caption{\textbf{Standard Deviation of Average Estimation Error with fixed time step $dt$}: We plot the relation between the total number of measurements with the standard deviation of the average estimation error $\sigma(J_k)$ over $100$ random realizations of the Hamiltonian. Horizontal axis shows the total number of measurements, where the total number of measurements is given by $8\times\{10^6, 10^7, 10^8, 10^9, 10^{10}\}$. The vertical axis shows the standard deviation $\sigma(J_k)$ for each Hankel matrix. }
%\label{fig:FixdtStdError}
%\end{figure}

The role of the total time $T$ is highlighted when we consider the second scenario where $T$ is fixed:  {we fix the total time evolution time $T$ used for constructing $\tilde{\mathbf{H}}_{4}$, and compare the estimation performance between $\tilde{\mathbf{H}}_{4}$ and $\tilde{\mathbf{H}}_{8}$. Then, the time step for $\tilde{\mathbf{H}}_{8}$ is chosen to be $dt'=\frac{7}{15}dt$. 
The result in Fig.~\ref{fig:FixT4vs8}} shows that in this case the larger Hankel matrix leads to larger errors although the time step $dt'$ satisfies the sampling theorem. In the presence of noise, 
the additional sampling points acquired mostly contribute to increase the noise, but do not convey much more information. In addition, 
a very small time step might lead to  larger errors, since it appears in the denominator of estimation equations [see, e.g., Eq.~(\ref{eq:Adt})].

\section{Discussion and conclusions}
\label{sec:conclusion}

Hamiltonian identification is a central task in the quest of constructing ever more complex quantum devices as well as characterizing and imaging quantum systems in biology and materials science. To access these systems at their nano-scale, we proposed to use a quantum probe that coherently couples to their dynamics. In this scenario, we re-analyzed Hamiltonian identification via the eigensystem realization algorithm (ERA) approach and provided a systematic algorithm to test identifiability by employing the Gr\"obner basis. Even more importantly from a practical point of view, we showed that analyzing these techniques yields bounds on the experimental resources required to estimate the Hamiltonian parameters, both in terms of the minimum coherence time required for Hamiltonian identification and for the overall total experimental time for the multi-parameter estimate. These bounds can guide experimentalists in implementing the most efficient Hamiltonian identification protocol. 
{We further numerically studied the estimation performance of ERA in the presence of noise. We found that the low-rank approximation for larger numbers of sampling points leads to more accurate estimation, even when the total number of measurements is kept fixed. This effects is however already at play for a small number of points above the minimum one, thus allowing one to keep the total evolution time short enough. When instead we fix the total evolution time as required to construct the smallest size  Hankel matrix, there is no longer an advantage in using a larger number of sampling points, as the smaller time step leads to larger estimation errors. These analyses quantitatively provide helpful insights for a practical experimental approach to  Hamiltonian identification based on ERA. }

In order to obtain exemplary analytical results for our Hamiltonian identification protocol, we considered simple models of spin chains coupled by one end to the quantum probe. While these models are less complex than what would be found in practical experimental scenarios, they allowed us to clarify an interesting relation between Hamiltonian identifiability by a quantum probe and quantum information propagation in a chain. Indeed, as Hamiltonian identification relies on  building a complete accessible set, the transport of information along the spin chain, in the form of spin-spin correlation, is a necessary condition. This result further imposes conditions on the time required for Hamiltonian identification: while in the cases we considered here these time bounds were consistent with the bounds directly imposed by ERA, it will be interesting to analyze in the future whether this result changes in the presence of disorder, when localization (either single particle or many-body) appears. 

We finally showed that by relaxing some of the assumptions on control constraints, by allowing for example collective control of the target system, can turn a previously non-identifiable system into an identifiable one. These results can contribute to make Hamiltonian identification more experimentally practical in many real-system scenarios.

\begin{acknowledgements}
A.S. thanks Daoyi Dong, Quntao Zhuang, and Can Gokler for fruitful discussions, and Chuteng Zhou and Yongbin Sun for their helpful numerical advice. This work was supported in part by the U.S. Army Research Office through Grant No. W911NF-11-1-0400 and W911NF-15-1-0548 and by the NSF PHY0551153. 
\end{acknowledgements}

\appendix
\section{Eigensystem Realization Algorithm}
We review the eigensystem realization algorithm and how it can be applied~\cite{Zhang14} for  Hamiltonian identification assisted by  single-probe measurement.  
\subsection{Construction of the state-space representation}
For an interacting $N$ spin-1/2 system, the Hamiltonian can be written as
\begin{equation}
H=\sum_{m=1}^{M}\theta_m S_m, 
\end{equation}
where $\theta_m\in\mathbb{R}\setminus\{0\}$ are the unknown parameters we want to determine, and $iS_m\in SU(2^{N})$. Let $\Gamma$ be the set of these Hermitian operators:
\begin{equation}
\Gamma=\{S_m|iS_m\in SU(2^{N})\}, m=1,2,\cdots, M\},
\label{operatorspace}
\end{equation}
with usually $M\ll 4^{N}-1$ due to the limitation in the number of spin couplings in the system. Let $G_0$ be the set of observables that we can  measure. The choice of $G_0$ is discussed in Sec.~\ref{sec: ERA}. We define the following iterative procedure:
\begin{equation}
G_j\equiv G_{j-1}\cup [G_{j-1},\Gamma],
\label{eq:iterative}
\end{equation}
where 
\begin{equation}
[G_{j-1},\Gamma]\equiv \{O_{i}|\text{tr}(O_{i}^{\dagger}[\eta,\gamma])\neq 0, \eta\in G_{j-1},\gamma\in \Gamma\}.
\end{equation}
Then, the finiteness in the dimension of $SU(2^{N})$ forces the iterative procedure to saturate, so that we can generate an accessible set $G$ of dimension $n\leq 4^{N}-1$:
\begin{equation}
G=\{O_{k}|i O_{k}\in SU(2^{N}), k=1,2,\cdots, n\}.
\label{eq:saturatedset}
\end{equation}
The physical meaning of $G$ was discussed in Sec.~\ref{sec: ERA}. 

The time evolution for each observable $O_k$ obeys  Heisenberg's equation:
\begin{equation}
\frac{dO_k}{dt}=i[H,O_k]=\sum_{l=1}^{n}\Big(\sum_{m=1}^{M}\theta_{m}V_{mkl}\Big)O_l,
\label{eq:Heisenberg}
\end{equation}
where
\begin{equation}
V_{mkl}=\text{Tr}(i[S_m, O_k]O_l)\in\mathbb{R}.
\end{equation}
Let $\rho_0$ be the initial state of the system, and let us define  $x_k=\text{Tr}(\rho_0 O_k)$. Eq.~(\ref{eq:Heisenberg}) can be written as
\begin{equation}
\frac{dx_k}{dt}=\sum_{l=1}^{n}\Big(\sum_{m=1}^{M}\theta_{m}V_{mkl}\Big)x_l
\label{eq:linear2}
\end{equation}
Defining a coherent vector $\mathbf{x}=(x_1,\cdots, x_{n})^{T}\in\mathbb{R}^{n}$, we can rewrite Eq.~ (\ref{eq:linear2}) into a compact form: 
\begin{equation}
\frac{d\mathbf{x}(t)}{dt}=\tilde{\mathbf{A}}\mathbf{x}(t),
\label{eq:linear3}
\end{equation}
where the \textit{system matrix} $\tilde{\mathbf{A}}\in \mathbb{R}^{n\times n}$ is a skew-symmetric matrix, i.e. $\tilde{\mathbf{A}}=-\tilde{\mathbf{A}}^{T}$. Let $y\in\mathbb{R}$ be the output data, which can be written in terms of the  output matrix $\mathbf{C}\in\mathbb{R}^{n}$ as
\begin{equation}
y(t)=\mathbf{C}\mathbf{x}(t).
\label{eq:output}
\end{equation}
 From Eq.~(\ref{eq:linear3}) and Eq.~(\ref{eq:output}),  a state-space representation can be constructed as the following:
\begin{equation}
\begin{split}
\frac{d\mathbf{x}(t)}{dt}&=\tilde{\mathbf{A}}\mathbf{x}(t)\\
y(t)&=\mathbf{C}\mathbf{x}(t).
\end{split}
\label{eq:statespace}
\end{equation}
In  discrete-time form, we have:
\begin{equation}
\begin{split}
\mathbf{x}(j+1)&=\mathbf{A}\mathbf{x}(j)\\
y(j)&=\mathbf{C}\mathbf{x}(j),
\end{split}
\end{equation}
where $\mathbf{x}(j)\equiv\mathbf{x}(j\Delta t), y(j)\equiv y(j\Delta t)$ and 
\begin{equation}
\mathbf{A}=e^{\tilde{\mathbf{A}}\Delta t}.
\end{equation}
Since any matrix exponential is a nonsingular matrix, we have:
\begin{equation}
\text{rank}(\mathbf{A})=n.
\label{eq:rankA}
\end{equation}
From Eq.~(\ref{eq:statespace}) we can obtain the  transfer function $T(s)=\mathbf{C}(sI_n-\tilde{\mathbf{A}})^{-1}\mathbf{x}(0)$, and $[\tilde{\mathbf{A}},\mathbf{C},\mathbf{x}(0)]$ is called the \textit{realization} of $T(s)$.

\subsection{Realization Theory and Hankel matrix}
\label{sec:Hankel}
The Hamiltonian identification algorithm~\cite{Zhang14} relies on  realization theory~\cite{Katayama06}. From the measurement data, we can construct the following Hankel matrix:
\begin{equation}
\mathbf{H}_{rs}(0)=
\begin{pmatrix}
y(0)&y(1)&\cdots& y(s-1)\\
y(1)&y(2)&\cdots& y(s)\\
\vdots&\vdots&\ddots&\vdots\\
y(r-1)&y(r)&\cdots& y(r+s-2)
\end{pmatrix},
\label{eq:Hankel0}
\end{equation} 
where we take $r,s\ge n$.  In order to obtain the transfer function with the true model order $n$, we need $r,s\ge n$ because the rank of the Hankel matrix is equal to the order of the transfer function. Suppose that $r,s<n$, meaning that one takes fewer observations. In general, the rank of a $r\times s$ matrix cannot be greater than either $r$ or $s$; therefore, we have:  
\begin{align*}
\text{rank}(\mathbf{H}_{rs}(0))\le \text{min}(r,s)<n.
\end{align*}
This means that a transfer function constructed from this smaller Hankel matrix would not have the true model order $n$. Therefore, $r$ and $s$ must satisfy $r,s\ge n$, and this is a necessary condition for ERA.

The Hankel matrix can be decomposed into
\begin{equation}
\mathbf{H}_{rs}(0)
=\mathcal{O}_{r}\mathcal{C}_{s},
\label{eq:Hankel1}
\end{equation}
where $\mathcal{O}_{r}\in\mathbb{R}^{rn\times n}$ and $\mathcal{C}_{s}\in\mathbb{R}^{n\times sn}$ are called \textit{observability} and \textit{controllability} matrix, respectively, with:
\begin{equation}
\begin{split}
\mathcal{O}_{r}&=\begin{pmatrix}
\mathbf{C}\\
\mathbf{C}\mathbf{A}\\
\vdots\\
\mathbf{C}\mathbf{A}^{r-1}
\end{pmatrix}\\
\mathcal{C}_{s}&=\begin{pmatrix}
\mathbf{x}(0)&\mathbf{A}\mathbf{x}(0)&\cdots&\mathbf{A}^{s-1}\mathbf{x}(0)
\end{pmatrix}.
\end{split}
\end{equation}
The singular value decomposition of $\mathbf{H}_{rs}(0)$ yields:
\begin{equation}
\mathbf{H}_{rs}(0)=\mathbf{U}
\begin{pmatrix}
\bf{\Sigma} &\mathbf{O}\\
\mathbf{O}&\mathbf{O}
\end{pmatrix}
\mathbf{V}^{T}
=
\begin{pmatrix}
\mathbf{U}_1&\mathbf{U}_2
\end{pmatrix}
\begin{pmatrix}
\bf{\Sigma} &\mathbf{O}\\
\mathbf{O}&\mathbf{O}
\end{pmatrix}
\begin{pmatrix}
\mathbf{V}_{1}^{T}\\
\mathbf{V}_{2}^{T}
\end{pmatrix}
\end{equation}
where $\mathbf{U}$ and $\mathbf{V}$ are  unitary matrices of  dimensions $rn\times rn$ and $sn\times sn$, respectively. Let $l\le n$ be the number of non-zero singular values of $\mathbf{H}_{rs}(0)$. $\bf{\Sigma}$ is the $l\times l$ diagonal matrix containing the non-zero singular values. Therefore, the observability and controllability matrices become:
\begin{equation}
\begin{split}
\mathcal{O}_{r}&=\mathbf{U}_1\bf{\Sigma}^{1/2}\\
\mathcal{C}_{s}&=\bf{\Sigma}^{1/2}\mathbf{V}_1^{T}.
\end{split}
\label{eq:C and O}
\end{equation}
By introducing the shifted-Hankel matrix:
\begin{equation}
\mathbf{H}_{rs}(1)=
\begin{pmatrix}
y(1)&y(2)&\cdots& y(s)\\
y(2)&y(3)&\cdots& y(s+1)\\
\vdots&\vdots&\ddots&\vdots\\
y(r)&y(r+1)&\cdots& y(r+s-1)
\end{pmatrix}=\mathcal{O}_{r}\mathbf{A}\mathcal{C}_{s},
\label{eq:Hankel1}
\end{equation} 
from Eq.~(\ref{eq:Hankel1}) and Eq.~(\ref{eq:C and O}), we can obtain the following \textit{new realization} of the transfer function: $[\tilde{\mathbf{A}}_{\text{est}},\mathbf{C}_{\text{est}},\mathbf{x}_{\text{est}}(0)]$ such that:
\begin{equation}
\begin{split}
\mathbf{x}_{\text{est}}(0)&=(\mathcal{O}_r)_{\text{first column}}\\
\mathbf{C}_{\text{est}}&=(\mathcal{C}_s)_{\text{first row}}\\
\tilde{\mathbf{A}}_{\text{est}}&=\frac{1}{\Delta t}\ln [\mathcal{O}_{r}^{-1}\mathbf{H}_{rs}(1)\mathcal{C}_{s}^{-1}].
\end{split}
\label{eq:Adt}
\end{equation}
We write the corresponding transfer fuction as
\begin{equation}
T_{\text{est}}(s)=\mathbf{C}_{\text{est}}(s I_n-\tilde{\mathbf{A}}_{\text{est}})^{-1}\mathbf{x}_{\text{est}}(0),
\label{eq:transferest}	
\end{equation}
and, in principle, $T(s)=T_{\text{est}}(s)$.
In order to obtain the new realization, the system is required to be both observable and controllable. Therefore, the controllability and observability matrix must satisfy: 
\begin{equation}
\begin{split}
\text{rank}(\mathcal{C}_s)&=n=\text{rank}(\mathbf{A})\\
\text{rank}(\mathcal{O}_r)&=n=\text{rank}(\mathbf{A}).
\end{split}
\label{eq:rankcondition}
\end{equation}
In turns, their ranks are determined by the Hankel matrix's rank, as required by the Sylvester inequality: for $\mathbf{P}\in\mathbb{R}^{m\times k}$, $\mathbf{Q}\in\mathbb{R}^{k\times n}$, 
\begin{align*}
p+q-k\le \text{rank}(\mathbf{PQ})\le \text{min}\{p,q\},
\end{align*}
where $p=\text{rank}(\mathbf{P})$ and $q=\text{rank}(\mathbf{Q})$.
From Eq.~(\ref{eq:Hankel1}) and Eq.~(\ref{eq:rankcondition}), the rank of the Hankel matrix must be:
\begin{equation}
\text{rank}(\mathbf{H}_{rs}(0))=n=\text{rank}(\mathbf{A}),
\end{equation}
which indicates that the minimum dimension of the Hankel matrix and the shifted Hankel matrix is $n\times n$. Therefore, all the output data $\{y(0),\dots,y(2n-1)\}$ need to be recorded, which means that we require at least $2n$ sampling points in order to obtain the new realization of the system and thus extract the unknown parameters. 
From Eq.~(\ref{eq:rankA}), the lower bound in the number of sampling points $\lambda_{\text{min}}$ is given by:
\begin{equation}
\lambda_{\text{min}}=2\text{rank}(\mathbf{A}).
\end{equation}

\section{Basic theory of Gr\"obner basis}
\label{sec:Groebner}
In this section, we review the basic theory of the Gr\"obner basis introduced in~\cite{Cox15, Cox04, Froeberg97, Becker93}.
\subsection{Monomial, polynomial, and monomial ordering}
\label{sec:monopoly}
Let $\mathbb{Z}_{\geq0}$ be the set of all nonnegative integers. A monomial in $z_1,\cdots, z_n$ is the product $z_1^{\alpha_1}\cdots z_n^{\alpha_n}$,
where $\alpha_1,\cdots, \alpha_n\in\mathbb{Z}_{\geq0}$. For simplicity, let us introduce the vectors $z=(z_1,\cdots, z_n)\in k^{n}$ and $\alpha=(\alpha_1,\cdots,\alpha_n)\in\mathbb{Z}_{\ge 0}^{n}$. Then, we write monomials as $z^{\alpha}\equiv z_1^{\alpha_1}\cdots z_n^{\alpha_n}$ and the monomial degree is 
\begin{equation}
|\alpha|=\sum_{k=1}^{n}\alpha_k.
\end{equation}
A polynomial $f\in k[z_1,\cdots, z_n]$  is a finite linear combination of the monomials with coefficients in a field $k$:
\begin{equation}
f=\sum_{\alpha}c_{\alpha}z^{\alpha}, \ \ c_{\alpha}\in k.
\end{equation}

Monomial ordering is an important ingredient in all algorithms developed in  commutative algebra. Let us introduce the so-called \text{lexicographic order} (lex) that we adapt to the Hamiltonian identification problem. Suppose  we have $\alpha=(\alpha_1,\cdots,\alpha_n)\in\mathbb{Z}_{\geq0}^{n}$ and $\beta=(\beta_1,\cdots,\beta_n)\in\mathbb{Z}_{\geq0}^{n}$. If the leftmost nonzero entry of $\alpha-\beta\in\mathbb{Z}^{n}$ is positive, we write $\alpha\succ_{lex}\beta$ or $z^{\alpha}\succ_{lex}z^{\beta}$. For each variable $z_1,\cdots,z_n$, the variables are ordered in the following way according to the lex ordering: $z_1\succ_{lex} z_2\succ_{lex}\cdots\succ_{lex}z_n$. By fixing the monomial ordering $\succ$, we can define the following terms.
\begin{enumerate}
\item{The \textbf{multidegree} of a polynomial $f$ is : $\text{multideg}(f)=\text{max}(\alpha\in\mathbb{Z}_{\geq0}^{n}|c_{\alpha\neq0})$ with respect to $\succ$.}
\item{The \textbf{leading coefficient} of a polynomial $f$ is: $\text{LC}(f)=c_{\text{multideg}(f)}\in k$.}
\item{The \textbf{leading monomial} of $f$ is $\text{LM}(f)=x^{\text{multideg}(f)}.$}
\item{The \textbf{leading term} is $\text{LT}(f)=\text{LC}(f)\cdot\text{LM}(f)$.}
\end{enumerate}

\subsection{Ideals and affine variety}
\label{sec: Ideals}
Let $k$ be a commutative ring. A subset $I\subseteq k$ is called an ideal if it satisfies the following conditions:
\begin{enumerate}
\item{$0\in I$.}
\item{If $f,g\in I$, then $f+g\in I$.}
\item{If $f\in I$ and $h\in k$, then $hf\in I$.}
\end{enumerate}
We are in particular interested in \textit{polynomial ideals}. 
We denote the set of all polynomials in $z_1,\cdots, z_n$ with coefficients on a field $k$ by  $k[z_1,\cdots, z_n]$.
%Let $f_1,\cdots,f_p$ be polynomials on the field $k$ in $x_1,\cdots, x_n$
Then, a subset $\mathcal{I}=\langle f_1,\cdots, f_p\rangle \subseteq k[z_1,\dots,z_n]$ such that:
\begin{equation}
\mathcal{I}=\langle f_1,\cdots,f_p\rangle=\Big\{\sum_{k=1}^{p}h_kf_k\Big|h_1,\cdots,h_p\in k[z_1,\cdots, z_n]\Big\}
\end{equation}
is an ideal of $k[z_1,\cdots, z_n]$. We call $\mathcal{I}$ a \textit{polynomial ideal} generated by $f_1,\cdots, f_p$, and $f_1,\cdots, f_p$ are called the bases of the polynomial ideal $\mathcal{I}$.

The radical of $\mathcal{I}$ is defined by:
\begin{equation}
\sqrt{\mathcal{I}}=\{f\in k[z_1,\cdots, z_n]|f^{k}\in\mathcal{I} \ \text{for some integer}\  k\geq1\},
\end{equation}
and we always have $\mathcal{I}\subseteq\sqrt{\mathcal{I}}$. Particularly, when $\mathcal{I}=\sqrt{\mathcal{I}}$, $\mathcal{I}$ is called a radical ideal. 

Let $\mathbf{V}(f_1,\cdots, f_p)$ be the set of solutions of a system of polynomial equations, i.e.
\begin{equation}
\mathbf{V}(f_1,...,f_p)\!=\!\{(a_1,...,a_n)\!\in\! k^{n}\,|\,f_l(a_1,...,a_n)\!=\!0\}, 
\end{equation}
for $l=1,2,\cdots,p$. $\mathbf{V}(f_1,\cdots, f_p)$ is called the \textit{affine variety} defined by $f_1,\cdots,f_p$. If $\{f_1,\cdots, f_p\}$ and $\{g_1,\cdots, g_s\}$ are the bases of the same polynomial ideal $\mathcal{I}$, then $\mathbf{V}(f_1,\cdots, f_p)=\mathbf{V}(g_1,\cdots,g_s)$. Any polynomial ideal $\mathcal{I}$ always satisfies:
\begin{equation}
\mathbf{V}(\sqrt{\mathcal{I}})=\mathbf{V}(\mathcal{I}),
\end{equation}
and, particularly, if $k$ is an algebraically closed field $\mathbb{C}$, the affine variety and the radical ideal are in one-to-one correspondence.

\subsection{Gr\"obner basis}
For a polynomial ideal $\mathcal{I}\in k[z_1,\cdots, z_n]\setminus\{0\}$, fixing a monomial order $\succ$, we define the leading term of the ideal, $ 
\text{LT}(\mathcal{I})=\{\text{LT}(f)|\exists f\in\mathcal{I}\setminus\{0\}\}$, 
and we write the monomial ideal  generated by the elements of $\text{LT}(\mathcal{I})$ as $\langle \text{LT}(\mathcal{I})\rangle$. If $\mathcal{I}=\langle f_1,\cdots, f_p\rangle$, we always have $\text{LT}(f_k)\in\text{LT}(\mathcal{I})\subseteq \langle \text{LT}(\mathcal{I})\rangle$.
 
From the Hilbert basis theorem~\cite{Cox04}, every polynomial ideal $\mathcal{I}\setminus\{0\}$ has a finite generating set $\mathcal{G}(\mathcal{I})=\{g_1,\cdots, g_t\}$, which satisfies $\langle \text{LT}(\mathcal{I})\rangle=\langle \text{LT}(g_1),\cdots,\text{LT}(g_t)\rangle$. $\mathcal{G}(\mathcal{I})$ is called a \textit{Gr\"obner basis} for the polynomial ideal $\mathcal{I}$. Therefore, the Hilbert basis theorem suggests that every polynomial ideal has a corresponding Gr\"obner basis. By adding the following restrictions:
\begin{enumerate}
\item{every polynomial $g_j$ is monic. i.e. $\text{LC}(g_j)=1$;}
\item{for every set of two distict polynomial $g_j$ and $g_i$, $\text{LM}(g_j)$ is not divisible by $\text{LM}(g_i)$ for any $i\neq j$,}
\end{enumerate}
we can obtain a unique minimal basis. The Gr\"obner basis with these restrictions is called \textit{reduced Gr\"obner basis}, which is denoted by $\mathscr{G}(\mathcal{I})$.

\subsection{Buchberger's algorithm for constructing the Gr\"obner basis}
The Gr\"obner basis can be constructed by Buchberger's algorithm~\cite{Buchberger06}. Let $S_{i,j}$ be the S-polynomial of the pair $(f_i,f_j)$, which is defined as:
\begin{equation}
S_{i,j}=\frac{\text{LCM}[\text{LM}(f_i),\text{LM}(f_j)]}{\text{LM}(f_i)}f_i-\frac{\text{LCM}[\text{LM}(f_i),\text{LM}(f_j)]}{\text{LM}(f_j)}f_j,
\end{equation}
where $\text{LCM}[\text{LM}(f_i),\text{LM}(f_j)]$ denotes the least common multiple of $\text{LM}(f_i)$ and $\text{LM}(f_j)$. Let $\text{rem}(S_{i,j},G)$ be the remainder of dividing $S_{i,j}$ by all elements in $G$. Let us consider the ideal $\mathcal{I}\subset\mathbb{C}[z_1,\cdots,z_n]$ generated by $f_1,\cdots, f_p$. Buchberger's algorithm~\cite{Cox15} is given by:
 \begin{algorithmic}
 \STATE \textbf{INPUT}: $F=\{f_1,\cdots,f_p\}$
 \STATE \textbf{OUTPUT}: The Gr\"obner basis $\mathscr{G}(\mathcal{I})=\{g_1,\cdots,g_t\}$ for the ideal $\mathcal{I}$.
 
\STATE $G:=F$

\REPEAT 
\STATE $G':=G$
\FORALL{$\{f_i,f_j\}, i \neq j$ in $G'$}
\STATE $R:=\text{rem}(S_{i,j},G')$
\IF{$R\neq0$}
\STATE $G:=G\cup \{\text{rem}(S_{i,j},G')\}$
\ENDIF
\ENDFOR
\UNTIL{$G=G'$}
\STATE RETURN $G$
\end{algorithmic}

\subsection{Construction of radicals of zero-dimensional ideal}
Let us consider the following system of polynomial equations:
\begin{equation}
\begin{split}
f_1&(z_1,\cdots,z_n)=0\\
f_2&(z_1,\cdots,z_n)=0\\
&\vdots\\
f_p&(z_1,\cdots,z_n)=0
\end{split}
\label{eq:system1}
\end{equation}
and $f_1,\cdots,f_p\in\mathbb{C}[z_1,\cdots,z_n]$. Suppose that Eq.~(\ref{eq:system1}) has a finite set of solutions. Then, the polynomial ideal $\mathcal{I}$ generated by $f_1,\cdots,f_n$ is called zero-dimensional ideal. Here, let us introduce the procedure to construct the radical $\sqrt{\mathcal{I}}$. By Buchberger's algorithm and the definition of the reduced Gr\"obner basis, we can obtain the following reduced Gr\"obner basis:
\begin{equation}
\begin{split}
\mathscr{G}(\mathcal{I})=\{&g_1(z_1),\\
                           &g_{2,1}(z_1,z_2),\cdots,g_{2,v_{2}}(z_1,z_2),\\
                           &\vdots\\ 
                           &g_{M,1}(z_1,\cdots,z_M),\cdots,g_{M,v_M}(z_1,\cdots,z_M)\}. 
\end{split}
\end{equation}
and $\mathscr{G}(\mathcal{I})$ generates the same ideal. Let $h_j$ be an unique monic generator of the elimination ideal $\mathcal{I}\cap\mathbb{C}[z_j]$. Then, we can choose $h_j$ such that $h_j\in\mathscr{G}(\mathcal{I})\cap\mathbb{C}[z_j]$ by the elimination theorem. Let $\varphi_j(z_j)$ be $\varphi_j=h_j/\text{gcd}(h_j,\partial_{z_j}h_j)$, the radical of the zero-dimensional ideal $\mathcal{I}$ is given by:
\begin{equation}
\sqrt{\mathcal{I}}=\mathcal{I}+\langle \varphi_1,\cdots,\varphi_n\rangle.
\end{equation}
(See e.g., ~\cite{Becker93} for the proof). In particular, by Seidenberg's Lemma~\cite{Becker93}, when $\varphi_j=1$,  $\mathcal{I}$ is a radical zero-dimensional ideal, and the Gr\"obner basis for the radical zero-dimensional ideal has a special shape, as described by the shape lemma~\cite{Cox04}, such that:
\begin{equation}
\mathscr{G}=\{z_1^{\alpha}+q_1(z_1),\cdots,z_{n-1}+q_{n-1}(z_1),z_n+q_n(z_1)\},
\end{equation}
where $\alpha\in\mathbb{N}$ and $q_{j}(z_1)$ are the univariate polynomials in $z_1$ with degree $\text{deg}(q_j)<\alpha$. 

\subsection{Elimination theory}
Let $\mathcal{I}\subseteq k[z_1,\cdots, z_n]$ be a polynomial ideal. Let us define $\mathcal{I}_l$ by:
\begin{equation}
\mathcal{I}_l=\mathcal{I}\cap k[z_{l+1},\dots, z_n],
\end{equation}
and we call $\mathcal{I}_l$ the $l$-th elimination ideal. Fixing the lex order $z_1\succ_{lex}z_2\succ_{lex}\cdots\succ_{lex}z_n$, for every $l$, the Gr\"obner basis for the $l$-th elimination ideal is written by:
\begin{equation}
\mathcal{G}_l=\mathcal{G}\cap k[z_{l+1},\cdots, z_n],
\end{equation}
where $\mathcal{G}$ is the Gr\"obner basis for $\mathcal{I}$ (elimination theorem)~\cite{Cox15}.

By employing the elimination theorem, we can derive the shape of the reduced Gr\"obner basis in Sec.~\ref{sec:isingtr} and Sec.~\ref{sec:exchange}. Let us take $\mathbf{x}(0)=(1,0,\cdots,0)^{T}\in\mathbb{R}^{n+1}$, and $\mathbf{C}=\begin{pmatrix}1&0&\cdots&0\end{pmatrix}\in\mathbb{R}^{n+1}$. The system matrix $\tilde{\mathbf{A}}$ is an $(n+1)\times (n+1)$ skew-symmetric matrix with the only non-zero elements $\theta_k=(\tilde{\mathbf{A}})_{k+1,k}=-(\tilde{\mathbf{A}})_{k,k+1}$, where $k=1,2,\cdots,n$. Then, from Eq. (\ref{eq:identity}), we  obtain the following system of polynomial equations:
\begin{equation}
f_1(z_1,\cdots,z_{n})=\cdots=
f_{n}(z_1,\cdots,z_{n})=0,
\end{equation}
where $z_k=\theta_l^{2}\ (k,l=1,2,\cdots,n)$. We can construct a polynomial ideal
\begin{equation}
\mathcal{I}=\langle f_1,\cdots, f_n\rangle\in\mathbb{C}[z_1,\cdots, z_n].
\end{equation}
Note that for convenience we consider the polynomial ideal over the polynomial ring $\mathbb{C}[z_1,\cdots,z_n]$. In this case, we have found that there exists a proper choice for the pair $(k,l)$ such that the corresponding elimination ideal $\mathcal{I}_{l-1}$ has the basis  $z_k-c_k^2$ for $\exists c_k\in\mathbb{R}$, meaning that: 
\begin{equation}
z_k-c_k^2\in\mathcal{G}_{l-1}.
\end{equation}
From the elimination theorem, we have:
\begin{equation}
\begin{split}
\mathcal{G}_{l-1}&=\mathcal{G}\cap \mathbb{C}[z_{l},\cdots,z_n]\\
&=\mathcal{G}\cap(\mathbb{C}[z_{l}]\cup\mathbb{C}[z_{l+1},\cdots,z_n])\\
&=(\mathcal{G}\cap\mathbb{C}[z_l])\cup(\mathscr{G}\cap\mathbb{C}[z_{l+1},\cdots, z_n])\\
&=(\mathcal{G}\cap\mathbb{C}[z_l])\cup\mathcal{G}_l,
\end{split}
\end{equation}
which yields:
\begin{equation}
\mathcal{G}_l\subset \mathcal{G}_{l-1}.
\end{equation}
Therefore, we can inductively obtain: 
\begin{equation}
\mathcal{G}_{n-1}\subset\mathcal{G}_{n-2}\subset\cdots\subset\mathcal{G}_{2}\subset\mathcal{G}_{1}\subset\mathcal{G}.
\end{equation}

By the definition of the reduced Gr\"obner basis, $\mathscr{G}(\mathcal{I})$ has the shape:
\begin{equation}
\mathscr{G}(\mathcal{I})=\langle z_1-a_1,\cdots,z_n-a_n\rangle,
\end{equation}
where $a_k=c_l^{2}$. This tells us the fact that $\mathcal{I}$ is the maximal ideal of $\mathbb{C}[z_1,\cdots, z_n]$.

\subsection{Comments on efficiency of Gr\"obner basis}
The computation of Gr\"obner basis takes tremendously large complexity \cite{Cox15}. Let $F$ be a set of polynomials $\{f_1,\cdots, f_t\}$ in $z_{1},\cdots, z_{n}\in\mathbb{C}$, and let $d$ be the maximal multiple degree of the input polynomials, i.e. $d=\text{max}(\text{multideg}(f_{1}), \cdots, \text{multideg}(f_{t}))$. 
Suppose that $F$ generates a zero-dimensional ideal. Then, the complexity for computing the reduced Gr\"obner basis can be given by $d^{O(n)}$ \cite{Lakshman91}. Therefore, for a larger system, the Gr\"obner basis takes a tremendously long time due to its complexity. Efficiency improvement of computing Gr\"obner basis is a timely problem. For example, recently, Gritzmann and Sturmfels proposed the idea of dynamic alternation of the monomial ordering while the algorithm progresses \cite{Gritzmann93, Caboara14}. Therefore, we expect that the current research efforts on the development efficient computation method of Gr\"obner basis can definitely contribute to the reduction of the computation complexity in the Hamiltonian identification. 
We want to emphasize that the Gr\"obner basis approach is a fundamental and systematic way to solve the system of polynomial equations. More importantly, it is useful due to peculiar properties of its shape which can determine the solvability of the system of polynomial equations and hence the identifiability of the Hamiltonian. Therefore, learning the Hamiltonian identifiability by applying Gr\"obner basis is fundamentally essential and necessary.

\section{Examples of polynomials for identifiable Hamiltonians}
\label{sec:Polynomials}
In this section, we show the explicit polynomials for particular identifiable models.
\subsection{$N=3$ Ising model with transverse field}
\label{sec:PolyIsing}
For the Ising model with transverse field with $N=3$ spins, the Hamiltonian can be written as:
\begin{align*}
H=\sum_{k=1}^{3}\frac{\omega_{k}}{2}Z_{k}+\sum_{k=1}^{2}\frac{J_{k}}{2}X_{k}X_{k+1}.
\end{align*}
Let us choose $G_{0}=\{X_{1}\}$ and let the initial state of the probe be the eigenstate of $X_{1}$ and the rest of spins in the chain be the maximally mixed state. The system matrix $\tilde{\mathbf{A}}$ is
\begin{align*}
\tilde{\mathbf{A}}=
\begin{pmatrix}
0&-\omega_{1}&0&0&0&0\\
\omega_{1}&0&-J_{1}&0&0&0\\
0&J_{1}&0&-\omega_{2}&0&0\\
0&0&\omega_{2}&0&-J_{2}&0\\
0&0&0&J_{2}&0&-\omega_{3}\\
0&0&0&0&\omega_{3}&0
\end{pmatrix},\\
\end{align*}
and output matrix $\mathbf{C}$ is given as
\begin{align*}
\mathbf{C}=\begin{pmatrix}1&0&0&0&0&0\end{pmatrix},
\end{align*}
and the initial coherent vector is:
\begin{align*}
\mathbf{x}(0)&=(1,0,0,0,0,0)^{T}.
\end{align*}
Let us define $(z_1,z_2,z_3,z_4,z_5)=(\omega_{1}^{2},\omega_{2}^{2},\omega_{3}^{2},J_{1}^{2},J_{2}^{2})$.
Then, from Eq.~(\ref{eq:identity}), we can obtain the following form of the system of polynomial equations:
\begin{align*}
\begin{split}
&z_{1}z_{2}z_{3}=v_{1}\\
&z_1z_{2}+z_{1}z_{3}+z_{2}z_{3}+z_{3}z_{4}+z_{1}z_{5}+z_{4}z_{5}=v_2\\
&z_1+z_2+z_3+z_4+z_5=v_3\\
&z_2 z_3+z_3z_4+z_4z_5=v_4\\
&z_2+z_3+z_4+z_5=v_5,
\end{split}
\end{align*}
where $v_{k}>0~(k=1,\cdots,5)$. Then, the Gr\"obner basis takes the following form: $\mathscr{G}=\{z_{1}-a_{1}^2,z_{2}-a_{2}^2,z_{3}-a_{3}^2,z_{4}-a_{4}^2,z_{5}-a_{5}^2\}$,
where $\{a_1^2,a_2^2,a_3^2,a_4^2,a_5^2\}$ are given in Eq.~(\ref{eq:N3Ising}):
\begin{widetext}
\begin{equation}
\begin{split}
a_{1}^2&=v_{3}-v_{5}\\
a_{2}^2&=\frac{v_{2}-v_{4}}{v_{3}-v_{5}}+\frac{v_1+v_4(v_5-v_3)}{v_4-v_2+v_5(v_3-v_5)}\\
a_{3}^2&=\frac{v_1(v_4-v_2+v_5(v_5-v_3))}{(v_2-v_4)^{2}+v_3^2v_4-v_3v_5(v_2+v_4)+v_2v_5^2+v_1(v_5-v_3)}\\
a_{4}^2&=\frac{v_4-v_2}{v_3-v_5}\\
a_{5}^2&=-\frac{v_1}{v_2}+\frac{v_1-v_4(v_3-v_5)}{v_2-v_4-v_5(v_3-v_5)}+\frac{v_1(v_1(v_5-v_3)+v_4(-v_2+v_4+v_3(v_3-v_5)))}{v_2((v_2-v_4)^{2}-v_1v_3+v_3^{2}v_4+v_5(v_1-v_3(v_2+v_4))+v_2v_5^{2})},
\end{split}
\label{eq:N3Ising}
\end{equation}
\end{widetext}

\subsection{$N=4$ Exchange model without transverse field}
\label{sec:PolyExchange}
Next, let us consider the exchange model without transverse field with $N=4$ spins. The Hamiltonian can be written as:
\begin{align*}
H=\sum_{k=1}^{3}\frac{J_1}{2}(X_{k}X_{k+1}+Y_{k}Y_{k+1}).
\end{align*}
Let us take same observable set and initial state of spin chain in Sec.~\ref{sec:PolyIsing}. The system matrix $\tilde{\mathbf{A}}$ is
\begin{align*}
\tilde{\mathbf{A}}=
\begin{pmatrix}
0&-J_{1}&0&0\\
J_{1}&0&J_{2}&0\\
0&-J_{2}&0&-J_{3}\\
0&0&J_{3}&0
\end{pmatrix},\\
\end{align*}
and output matrix $\mathbf{C}$ is given as
\begin{align*}
\mathbf{C}=\begin{pmatrix}1&0&0&0\end{pmatrix},
\end{align*}
and the initial coherent vector is:
\begin{align*}
\mathbf{x}(0)&=(1,0,0,0)^{T}.
\end{align*}
Let us define $(z_1,z_2,z_3)=(J_{1}^{2},J_{2}^{2},J_{3}^{2})$.
Then, from Eq.~(\ref{eq:identity}), we can obtain the following form of the system of polynomial equations:
\begin{align*}
\begin{split}
&z_2+z_3=v_1\\
&z_1 z_3=v_2\\
&z_1+z_2+z_3=v_3\\
\end{split}
\end{align*}
where $v_{k}>0~(k=1,\cdots,5)$. Then, the Gr\"obner basis takes the following form: $\mathscr{G}=\{z_{1}-a_{1}^2,z_{2}-a_{2}^2,z_{3}-a_{3}^2\}$,
where $\{a_1^2,a_2^2,a_3^2\}$ are given in Eq.~(\ref{eq:N6Exchange}):
\begin{equation}
\begin{split}
a_1^2&=v_3-v_1\\
a_2^2&=\frac{v_1v_3-v_2-v_1^2}{v_3 - v1}\\
a_3^3&=\frac{v_2}{v_3 - v1}
\end{split}
\label{eq:N6Exchange}
\end{equation}
where $v_3>v_1$, $v_1(v_3-v_1)>v_2>0$.

\subsection{$N=2$ Exchange model with transverse field}
\label{sec:PolyExchangeTr}
Finally, let us consider the exchange model with $N=2$ spins with transverse field. The Hamiltonian can be written as:
\begin{align*}
H=\frac{\omega_{1}}{2}Z_{1}+\frac{\omega_{2}}{2}Z_{2}+\frac{J_{1}}{2}(X_1X_{2}+Y_1Y_{2}).
\end{align*}
In Sec.~\ref{sec:exchangetr}, we have discussed that the Hamiltonian becomes fully identifiable if we measure $X_{1}$ and $Y_{1}$ separately.
Let us always prepare the initial state of the spin probe to be the eigenstate of $\{X_{1}\}$ and the other spin to be the maximally mixed state. The system matrix is
\begin{align*}
\tilde{\mathbf{A}}=
\begin{pmatrix}
0&\omega_{1}&0&-J_1\\
-\omega_{1}&0&-J_1&0\\
0&J_{1}&0&\omega_{2}\\
J_{1}&0&-\omega_{2}&0
\end{pmatrix}.
\end{align*}
Here, the output matrix becomes 
\begin{align*}
\mathbf{C}=\begin{pmatrix}1&1&0&0\end{pmatrix}
\end{align*}
and the initial coherent vector is
\begin{align*}
\mathbf{x}(0)=(1,0,0,0)^{T}.
\end{align*}
Let us define $(z_{1},z_{2},z_{3})=(\omega_{1},\omega_{2},J_1^2)$. 
Then, from Eq.~(\ref{eq:identity}), we can obtain the following form of the system of polynomial equations:
\begin{align*}
\begin{split}
&-z_1z_2^2-z_2 z_3=v_1\\
&z_2^2+z_3=v_2\\
&z_1=v_3\\
&(z_1 z_2+z_3)^2=v_4\\
&z_1^2+z_2^2+2 z_3=v_5.
\end{split}
\end{align*}
Then, the Gr\"obner takes the following form: $\mathscr{G}=\{z_1-a_1,z_2-a_2,z_3-a_3^2\}$,
where $\{a_1,a_2,a_3^2\}$ are given in Eq.~(\ref{eq:N2ExcTr}):
\begin{equation}
\begin{split}
a_1&=v_3\\
a_2&=\frac{v_1 + 2 v_2 v_3 + v_3^3 - v_3 v_5}{v_2 + v_3^2 - v_5}\\
&=\frac{v_2^2 - v_2 v_3^2 - v_3^4 + v_4 - v_2 v_5 + v_3^2 v_5}{v_1 - v_3^3 + v_3 v_5}\\
&=\frac{2 v_1 (v_2 - v_5) + 
 v_3 (-v_2^2 + 2 v_2 v_3^2 + 2 v_3^4 - 3 v_4 - 3 v_3^2 v_5 + v_5^2)}{2 (v_3^4 + v_4 - v_3^2 v_5)}\\
a_3^2&=v_5-v_2-v_3^2,
\end{split}
\label{eq:N2ExcTr}
\end{equation}
where $v_2+v_3^2-v_5\neq0$, $v_1-v_3^2+v_3v_5\neq0$, $v_3^4+v_4-v_3^2v_5\neq0$ and $v_5-v_2-v_3^2>0$. Here $a_1$ and $a_2$ are the nonzero real numbers. Also note that $\{v_1,v_2,v_3,v_4,v_5\}$ satisfy the following simultaneous identity:
\begin{widetext}
\begin{align*}
&v_2^4 - 4 v_2^3 v_5 - 2 v_2^2 (v_4 + (v_3^2 - 3 v_5) v_5) + (v_4 - v_5^2)^2 + 
 v_3^4 (-4 v_4 + v_5^2) + v_3^2 (6 v_4 v_5 - 2 v_5^3) + 
 4 v_2 (v_5 (v_4 - v_5^2) + v_3^2 (-2 v_4 + v_5^2))=0\\
&-v_2^2 + 2v_1 v_3 + v_4 + 2 v_2 v_5 + (v_3^2 - v_5) v_5=0\\
&2 v_1 (-v_4 + (v_2 - v_5)^2) - v_3 (4 (2 v_2 + v_3^2) v_4 + (v_2^2 - 5 v_4) v_5 - (2 v_2 + v_3^2) v_5^2 + v_5^3)=0\\
& v_1^2 - v_4(2 v_2 + v_3^2 - v_5)=0.
 \end{align*}
\end{widetext}

\bibliography{Biblio}

\begin{thebibliography}{49}%
\makeatletter
\providecommand \@ifxundefined [1]{%
 \@ifx{#1\undefined}
}%
\providecommand \@ifnum [1]{%
 \ifnum #1\expandafter \@firstoftwo
 \else \expandafter \@secondoftwo
 \fi
}%
\providecommand \@ifx [1]{%
 \ifx #1\expandafter \@firstoftwo
 \else \expandafter \@secondoftwo
 \fi
}%
\providecommand \natexlab [1]{#1}%
\providecommand \enquote  [1]{``#1''}%
\providecommand \bibnamefont  [1]{#1}%
\providecommand \bibfnamefont [1]{#1}%
\providecommand \citenamefont [1]{#1}%
\providecommand \href@noop [0]{\@secondoftwo}%
\providecommand \href [0]{\begingroup \@sanitize@url \@href}%
\providecommand \@href[1]{\@@startlink{#1}\@@href}%
\providecommand \@@href[1]{\endgroup#1\@@endlink}%
\providecommand \@sanitize@url [0]{\catcode `\\12\catcode `\$12\catcode
  `\&12\catcode `\#12\catcode `\^12\catcode `\_12\catcode `\%12\relax}%
\providecommand \@@startlink[1]{}%
\providecommand \@@endlink[0]{}%
\providecommand \url  [0]{\begingroup\@sanitize@url \@url }%
\providecommand \@url [1]{\endgroup\@href {#1}{\urlprefix }}%
\providecommand \urlprefix  [0]{URL }%
\providecommand \Eprint [0]{\href }%
\providecommand \doibase [0]{http://dx.doi.org/}%
\providecommand \selectlanguage [0]{\@gobble}%
\providecommand \bibinfo  [0]{\@secondoftwo}%
\providecommand \bibfield  [0]{\@secondoftwo}%
\providecommand \translation [1]{[#1]}%
\providecommand \BibitemOpen [0]{}%
\providecommand \bibitemStop [0]{}%
\providecommand \bibitemNoStop [0]{.\EOS\space}%
\providecommand \EOS [0]{\spacefactor3000\relax}%
\providecommand \BibitemShut  [1]{\csname bibitem#1\endcsname}%
\let\auto@bib@innerbib\@empty
%</preamble>
\bibitem [{\citenamefont {Ajoy}\ \emph {et~al.}()\citenamefont {Ajoy},
  \citenamefont {Liu}, \citenamefont {Saha}, \citenamefont {Marseglia},
  \citenamefont {Jaskula}, \citenamefont {Bissbort},\ and\ \citenamefont
  {Cappellaro}}]{Ajoy16x}%
  \BibitemOpen
  \bibfield  {author} {\bibinfo {author} {\bibfnamefont {A.}~\bibnamefont
  {Ajoy}}, \bibinfo {author} {\bibfnamefont {Y.~X.}\ \bibnamefont {Liu}},
  \bibinfo {author} {\bibfnamefont {K.}~\bibnamefont {Saha}}, \bibinfo {author}
  {\bibfnamefont {L.}~\bibnamefont {Marseglia}}, \bibinfo {author}
  {\bibfnamefont {J.-C.}\ \bibnamefont {Jaskula}}, \bibinfo {author}
  {\bibfnamefont {U.}~\bibnamefont {Bissbort}}, \ and\ \bibinfo {author}
  {\bibfnamefont {P.}~\bibnamefont {Cappellaro}},\ }\href
  {http://arxiv.org/abs/1604.01677} {\bibinfo  {journal} {arXiv:1604.01677}\
  }\BibitemShut {NoStop}%
\bibitem [{\citenamefont {Ajoy}\ \emph {et~al.}(2015)\citenamefont {Ajoy},
  \citenamefont {Bissbort}, \citenamefont {Lukin}, \citenamefont {Walsworth},\
  and\ \citenamefont {Cappellaro1}}]{Ajoy15}%
  \BibitemOpen
\bibfield  {journal} {  }\bibfield  {author} {\bibinfo {author} {\bibfnamefont
  {A.}~\bibnamefont {Ajoy}}, \bibinfo {author} {\bibfnamefont {U.}~\bibnamefont
  {Bissbort}}, \bibinfo {author} {\bibfnamefont {M.~D.}\ \bibnamefont {Lukin}},
  \bibinfo {author} {\bibfnamefont {R.~L.}\ \bibnamefont {Walsworth}}, \ and\
  \bibinfo {author} {\bibfnamefont {P.}~\bibnamefont {Cappellaro1}},\ }\href
  {http://link.aps.org/doi/10.1103/PhysRevX.5.011001} {\bibfield  {journal}
  {\bibinfo  {journal} {Phys. Rev. X}\ }\textbf {\bibinfo {volume} {5}},\
  \bibinfo {pages} {011001} (\bibinfo {year} {2015})}\BibitemShut {NoStop}%
\bibitem [{\citenamefont {Lovchinsky}\ \emph {et~al.}(2016)\citenamefont
  {Lovchinsky}, \citenamefont {Sushkov}, \citenamefont {Urbach}, \citenamefont
  {de~Leon}, \citenamefont {Choi}, \citenamefont {De~Greve}, \citenamefont
  {Evans}, \citenamefont {Gertner}, \citenamefont {Bersin}, \citenamefont
  {M{\"u}ller}, \citenamefont {McGuinness}, \citenamefont {Jelezko},
  \citenamefont {Walsworth}, \citenamefont {Park},\ and\ \citenamefont
  {Lukin}}]{Lovchinsky16}%
  \BibitemOpen
  \bibfield  {author} {\bibinfo {author} {\bibfnamefont {I.}~\bibnamefont
  {Lovchinsky}}, \bibinfo {author} {\bibfnamefont {A.~O.}\ \bibnamefont
  {Sushkov}}, \bibinfo {author} {\bibfnamefont {E.}~\bibnamefont {Urbach}},
  \bibinfo {author} {\bibfnamefont {N.~P.}\ \bibnamefont {de~Leon}}, \bibinfo
  {author} {\bibfnamefont {S.}~\bibnamefont {Choi}}, \bibinfo {author}
  {\bibfnamefont {K.}~\bibnamefont {De~Greve}}, \bibinfo {author}
  {\bibfnamefont {R.}~\bibnamefont {Evans}}, \bibinfo {author} {\bibfnamefont
  {R.}~\bibnamefont {Gertner}}, \bibinfo {author} {\bibfnamefont
  {E.}~\bibnamefont {Bersin}}, \bibinfo {author} {\bibfnamefont
  {C.}~\bibnamefont {M{\"u}ller}}, \bibinfo {author} {\bibfnamefont
  {L.}~\bibnamefont {McGuinness}}, \bibinfo {author} {\bibfnamefont
  {F.}~\bibnamefont {Jelezko}}, \bibinfo {author} {\bibfnamefont {R.~L.}\
  \bibnamefont {Walsworth}}, \bibinfo {author} {\bibfnamefont {H.}~\bibnamefont
  {Park}}, \ and\ \bibinfo {author} {\bibfnamefont {M.~D.}\ \bibnamefont
  {Lukin}},\ }\href {\doibase 10.1126/science.aad8022} {\bibfield  {journal}
  {\bibinfo  {journal} {Science}\ }\textbf {\bibinfo {volume} {351}},\ \bibinfo
  {pages} {836} (\bibinfo {year} {2016})}\BibitemShut {NoStop}%
\bibitem [{\citenamefont {Cooper}\ \emph {et~al.}(2014)\citenamefont {Cooper},
  \citenamefont {Magesan}, \citenamefont {Yum},\ and\ \citenamefont
  {Cappellaro}}]{Cooper14}%
  \BibitemOpen
  \bibfield  {author} {\bibinfo {author} {\bibfnamefont {A.}~\bibnamefont
  {Cooper}}, \bibinfo {author} {\bibfnamefont {E.}~\bibnamefont {Magesan}},
  \bibinfo {author} {\bibfnamefont {H.}~\bibnamefont {Yum}}, \ and\ \bibinfo
  {author} {\bibfnamefont {P.}~\bibnamefont {Cappellaro}},\ }\href
  {http://dx.doi.org/10.1038/ncomms4141} {\bibfield  {journal} {\bibinfo
  {journal} {Nat. Commun.}\ }\textbf {\bibinfo {volume} {5}},\ \bibinfo {pages}
  {3141} (\bibinfo {year} {2014})}\BibitemShut {NoStop}%
\bibitem [{\citenamefont {Barry}\ \emph {et~al.}(2016)\citenamefont {Barry},
  \citenamefont {Turner}, \citenamefont {Schloss}, \citenamefont {Glenn},
  \citenamefont {Song}, \citenamefont {Lukin}, \citenamefont {Park},\ and\
  \citenamefont {Walsworth}}]{Barry16}%
  \BibitemOpen
  \bibfield  {author} {\bibinfo {author} {\bibfnamefont {J.~F.}\ \bibnamefont
  {Barry}}, \bibinfo {author} {\bibfnamefont {M.~J.}\ \bibnamefont {Turner}},
  \bibinfo {author} {\bibfnamefont {J.~M.}\ \bibnamefont {Schloss}}, \bibinfo
  {author} {\bibfnamefont {D.~R.}\ \bibnamefont {Glenn}}, \bibinfo {author}
  {\bibfnamefont {Y.}~\bibnamefont {Song}}, \bibinfo {author} {\bibfnamefont
  {M.~D.}\ \bibnamefont {Lukin}}, \bibinfo {author} {\bibfnamefont
  {H.}~\bibnamefont {Park}}, \ and\ \bibinfo {author} {\bibfnamefont {R.~L.}\
  \bibnamefont {Walsworth}},\ }\href {https://arxiv.org/pdf/1602.01056v1.pdf}
  {\bibfield  {journal} {\bibinfo  {journal} {Proc. Natl. Acad. Sci. USA}\
  }\textbf {\bibinfo {volume} {113}},\ \bibinfo {pages} {14133} (\bibinfo
  {year} {2016})}\BibitemShut {NoStop}%
\bibitem [{\citenamefont {van~der Sar}\ \emph {et~al.}(2014)\citenamefont
  {van~der Sar}, \citenamefont {Casola}, \citenamefont {Walsworth},\ and\
  \citenamefont {Yacoby}}]{Sar14}%
  \BibitemOpen
  \bibfield  {author} {\bibinfo {author} {\bibfnamefont {T.}~\bibnamefont
  {van~der Sar}}, \bibinfo {author} {\bibfnamefont {F.}~\bibnamefont {Casola}},
  \bibinfo {author} {\bibfnamefont {R.}~\bibnamefont {Walsworth}}, \ and\
  \bibinfo {author} {\bibfnamefont {A.}~\bibnamefont {Yacoby}},\ }\href
  {\doibase 10.1038/ncomms8886} {\bibfield  {journal} {\bibinfo  {journal}
  {Nat. Commun.}\ }\textbf {\bibinfo {volume} {6}},\ \bibinfo {pages} {7886}
  (\bibinfo {year} {2014})}\BibitemShut {NoStop}%
\bibitem [{\citenamefont {Wolfe}\ \emph {et~al.}(2016)\citenamefont {Wolfe},
  \citenamefont {Manuilov}, \citenamefont {Purser}, \citenamefont
  {Teeling-Smith}, \citenamefont {Dubs}, \citenamefont {Hammel},\ and\
  \citenamefont {Bhallamudi}}]{Wolfe16}%
  \BibitemOpen
  \bibfield  {author} {\bibinfo {author} {\bibfnamefont {C.~S.}\ \bibnamefont
  {Wolfe}}, \bibinfo {author} {\bibfnamefont {S.~A.}\ \bibnamefont {Manuilov}},
  \bibinfo {author} {\bibfnamefont {C.~M.}\ \bibnamefont {Purser}}, \bibinfo
  {author} {\bibfnamefont {R.}~\bibnamefont {Teeling-Smith}}, \bibinfo {author}
  {\bibfnamefont {C.}~\bibnamefont {Dubs}}, \bibinfo {author} {\bibfnamefont
  {P.~C.}\ \bibnamefont {Hammel}}, \ and\ \bibinfo {author} {\bibfnamefont
  {V.~P.}\ \bibnamefont {Bhallamudi}},\ }\href {\doibase 10.1063/1.4953108}
  {\bibfield  {journal} {\bibinfo  {journal} {App. Phys. Lett}\ }\textbf
  {\bibinfo {volume} {108}},\ \bibinfo {pages} {232409} (\bibinfo {year}
  {2016})}\BibitemShut {NoStop}%
\bibitem [{\citenamefont {Burgarth}\ \emph {et~al.}(2009)\citenamefont
  {Burgarth}, \citenamefont {Maruyama},\ and\ \citenamefont
  {Nori}}]{Burgarth09a}%
  \BibitemOpen
  \bibfield  {author} {\bibinfo {author} {\bibfnamefont {D.}~\bibnamefont
  {Burgarth}}, \bibinfo {author} {\bibfnamefont {K.}~\bibnamefont {Maruyama}},
  \ and\ \bibinfo {author} {\bibfnamefont {F.}~\bibnamefont {Nori}},\ }\href
  {\doibase 10.1103/PhysRevA.79.020305} {\bibfield  {journal} {\bibinfo
  {journal} {Phys. Rev. A}\ }\textbf {\bibinfo {volume} {79}},\ \bibinfo
  {pages} {020305(R)} (\bibinfo {year} {2009})}\BibitemShut {NoStop}%
\bibitem [{\citenamefont {Burgarth}\ and\ \citenamefont
  {Maruyama}(2009)}]{Burgarth09b}%
  \BibitemOpen
  \bibfield  {author} {\bibinfo {author} {\bibfnamefont {D.}~\bibnamefont
  {Burgarth}}\ and\ \bibinfo {author} {\bibfnamefont {K.}~\bibnamefont
  {Maruyama}},\ }\href {http://stacks.iop.org/1367-2630/11/i=10/a=103019}
  {\bibfield  {journal} {\bibinfo  {journal} {New J. Phys.}\ }\textbf {\bibinfo
  {volume} {11}},\ \bibinfo {pages} {103019} (\bibinfo {year}
  {2009})}\BibitemShut {NoStop}%
\bibitem [{\citenamefont {Franco}\ \emph {et~al.}(2009)\citenamefont {Franco},
  \citenamefont {Paternostro},\ and\ \citenamefont {Kim}}]{Franco09}%
  \BibitemOpen
  \bibfield  {author} {\bibinfo {author} {\bibfnamefont {C.~D.}\ \bibnamefont
  {Franco}}, \bibinfo {author} {\bibfnamefont {M.}~\bibnamefont {Paternostro}},
  \ and\ \bibinfo {author} {\bibfnamefont {M.~S.}\ \bibnamefont {Kim}},\ }\href
  {http://journals.aps.org/prl/pdf/10.1103/PhysRevLett.102.187203} {\bibfield
  {journal} {\bibinfo  {journal} {Phys. Rev. Lett.}\ }\textbf {\bibinfo
  {volume} {102}},\ \bibinfo {pages} {187203} (\bibinfo {year}
  {2009})}\BibitemShut {NoStop}%
\bibitem [{\citenamefont {Wang}\ \emph {et~al.}()\citenamefont {Wang},
  \citenamefont {Dong}, \citenamefont {Qi}, \citenamefont {Zhang},
  \citenamefont {Petersen},\ and\ \citenamefont {Yonezawa}}]{Wang16}%
  \BibitemOpen
  \bibfield  {author} {\bibinfo {author} {\bibfnamefont {Y.}~\bibnamefont
  {Wang}}, \bibinfo {author} {\bibfnamefont {D.}~\bibnamefont {Dong}}, \bibinfo
  {author} {\bibfnamefont {B.}~\bibnamefont {Qi}}, \bibinfo {author}
  {\bibfnamefont {J.}~\bibnamefont {Zhang}}, \bibinfo {author} {\bibfnamefont
  {I.~R.}\ \bibnamefont {Petersen}}, \ and\ \bibinfo {author} {\bibfnamefont
  {H.}~\bibnamefont {Yonezawa}},\ }\href {https://arxiv.org/pdf/1610.08841.pdf}
  {\bibinfo  {journal} {arXiv:1610.08841}\ }\BibitemShut {NoStop}%
\bibitem [{\citenamefont {Granade}\ \emph {et~al.}(2012)\citenamefont
  {Granade}, \citenamefont {Ferrie}, \citenamefont {Wiebe},\ and\ \citenamefont
  {Cory}}]{Granade12}%
  \BibitemOpen
\bibfield  {journal} {  }\bibfield  {author} {\bibinfo {author} {\bibfnamefont
  {C.~E.}\ \bibnamefont {Granade}}, \bibinfo {author} {\bibfnamefont
  {C.}~\bibnamefont {Ferrie}}, \bibinfo {author} {\bibfnamefont
  {N.}~\bibnamefont {Wiebe}}, \ and\ \bibinfo {author} {\bibfnamefont {D.~G.}\
  \bibnamefont {Cory}},\ }\href
  {http://iopscience.iop.org/article/10.1088/1367-2630/14/10/103013/pdf}
  {\bibfield  {journal} {\bibinfo  {journal} {New J. Phys.}\ }\textbf {\bibinfo
  {volume} {14}},\ \bibinfo {pages} {103013} (\bibinfo {year}
  {2012})}\BibitemShut {NoStop}%
\bibitem [{\citenamefont {Schirmer}\ and\ \citenamefont
  {Langbein}(2015)}]{Schirmer15}%
  \BibitemOpen
  \bibfield  {author} {\bibinfo {author} {\bibfnamefont {S.~G.}\ \bibnamefont
  {Schirmer}}\ and\ \bibinfo {author} {\bibfnamefont {F.~C.}\ \bibnamefont
  {Langbein}},\ }\href {\doibase 10.1103/PhysRevA.91.022125} {\bibfield
  {journal} {\bibinfo  {journal} {Phys. Rev. A}\ }\textbf {\bibinfo {volume}
  {91}},\ \bibinfo {pages} {022125} (\bibinfo {year} {2015})}\BibitemShut
  {NoStop}%
\bibitem [{\citenamefont {Sergeevich}\ \emph {et~al.}(2011)\citenamefont
  {Sergeevich}, \citenamefont {Chandran}, \citenamefont {Combes}, \citenamefont
  {Bartlett},\ and\ \citenamefont {Wiseman}}]{Sergeevich11}%
  \BibitemOpen
  \bibfield  {author} {\bibinfo {author} {\bibfnamefont {A.}~\bibnamefont
  {Sergeevich}}, \bibinfo {author} {\bibfnamefont {A.}~\bibnamefont
  {Chandran}}, \bibinfo {author} {\bibfnamefont {J.}~\bibnamefont {Combes}},
  \bibinfo {author} {\bibfnamefont {S.~D.}\ \bibnamefont {Bartlett}}, \ and\
  \bibinfo {author} {\bibfnamefont {H.~M.}\ \bibnamefont {Wiseman}},\ }\href
  {\doibase 10.1103/PhysRevA.84.052315} {\bibfield  {journal} {\bibinfo
  {journal} {Phys. Rev. A}\ }\textbf {\bibinfo {volume} {84}},\ \bibinfo
  {pages} {052315} (\bibinfo {year} {2011})}\BibitemShut {NoStop}%
\bibitem [{\citenamefont {Shabani}\ \emph {et~al.}(2011)\citenamefont
  {Shabani}, \citenamefont {Mohseni}, \citenamefont {Lloyd}, \citenamefont
  {Kosut},\ and\ \citenamefont {Rabitz}}]{Shabini11}%
  \BibitemOpen
  \bibfield  {author} {\bibinfo {author} {\bibfnamefont {A.}~\bibnamefont
  {Shabani}}, \bibinfo {author} {\bibfnamefont {M.}~\bibnamefont {Mohseni}},
  \bibinfo {author} {\bibfnamefont {S.}~\bibnamefont {Lloyd}}, \bibinfo
  {author} {\bibfnamefont {R.~L.}\ \bibnamefont {Kosut}}, \ and\ \bibinfo
  {author} {\bibfnamefont {H.}~\bibnamefont {Rabitz}},\ }\href
  {http://journals.aps.org/pra/pdf/10.1103/PhysRevA.84.012107} {\bibfield
  {journal} {\bibinfo  {journal} {Phys. Rev. A}\ }\textbf {\bibinfo {volume}
  {84}},\ \bibinfo {pages} {012107} (\bibinfo {year} {2011})}\BibitemShut
  {NoStop}%
\bibitem [{\citenamefont {Magesan}\ \emph {et~al.}(2013)\citenamefont
  {Magesan}, \citenamefont {Cooper},\ and\ \citenamefont
  {Cappellaro}}]{Magesan13c}%
  \BibitemOpen
  \bibfield  {author} {\bibinfo {author} {\bibfnamefont {E.}~\bibnamefont
  {Magesan}}, \bibinfo {author} {\bibfnamefont {A.}~\bibnamefont {Cooper}}, \
  and\ \bibinfo {author} {\bibfnamefont {P.}~\bibnamefont {Cappellaro}},\
  }\href {\doibase 10.1103/PhysRevA.88.062109} {\bibfield  {journal} {\bibinfo
  {journal} {Phys. Rev. A}\ }\textbf {\bibinfo {volume} {88}},\ \bibinfo
  {pages} {062109} (\bibinfo {year} {2013})}\BibitemShut {NoStop}%
\bibitem [{\citenamefont {Arai}\ \emph {et~al.}(2015)\citenamefont {Arai},
  \citenamefont {Belthangady}, \citenamefont {Zhang}, \citenamefont {Bar-Gill},
  \citenamefont {DeVience}, \citenamefont {Cappellaro}, \citenamefont
  {Yacoby},\ and\ \citenamefont {Walsworth}}]{Arai15}%
  \BibitemOpen
  \bibfield  {author} {\bibinfo {author} {\bibfnamefont {K.}~\bibnamefont
  {Arai}}, \bibinfo {author} {\bibfnamefont {C.}~\bibnamefont {Belthangady}},
  \bibinfo {author} {\bibfnamefont {H.}~\bibnamefont {Zhang}}, \bibinfo
  {author} {\bibfnamefont {N.}~\bibnamefont {Bar-Gill}}, \bibinfo {author}
  {\bibfnamefont {S.}~\bibnamefont {DeVience}}, \bibinfo {author}
  {\bibfnamefont {P.}~\bibnamefont {Cappellaro}}, \bibinfo {author}
  {\bibfnamefont {A.}~\bibnamefont {Yacoby}}, \ and\ \bibinfo {author}
  {\bibfnamefont {R.}~\bibnamefont {Walsworth}},\ }\href
  {http://dx.doi.org/10.1038/nnano.2015.171} {\bibfield  {journal} {\bibinfo
  {journal} {Nat. Nanotech.}\ }\textbf {\bibinfo {volume} {10}},\ \bibinfo
  {pages} {859} (\bibinfo {year} {2015})}\BibitemShut {NoStop}%
\bibitem [{\citenamefont {Zhang}\ and\ \citenamefont
  {Sarovar}(2014)}]{Zhang14}%
  \BibitemOpen
  \bibfield  {author} {\bibinfo {author} {\bibfnamefont {J.}~\bibnamefont
  {Zhang}}\ and\ \bibinfo {author} {\bibfnamefont {M.}~\bibnamefont
  {Sarovar}},\ }\href {\doibase 10.1103/PhysRevLett.113.080401} {\bibfield
  {journal} {\bibinfo  {journal} {Phys. Rev. Lett.}\ }\textbf {\bibinfo
  {volume} {113}},\ \bibinfo {pages} {080401} (\bibinfo {year}
  {2014})}\BibitemShut {NoStop}%
\bibitem [{\citenamefont {Zhang}\ and\ \citenamefont
  {Sarovar}(2015)}]{Zhang15}%
  \BibitemOpen
  \bibfield  {author} {\bibinfo {author} {\bibfnamefont {J.}~\bibnamefont
  {Zhang}}\ and\ \bibinfo {author} {\bibfnamefont {M.}~\bibnamefont
  {Sarovar}},\ }\href {\doibase 10.1103/PhysRevA.91.052121} {\bibfield
  {journal} {\bibinfo  {journal} {Phys. Rev. A}\ }\textbf {\bibinfo {volume}
  {91}},\ \bibinfo {pages} {052121} (\bibinfo {year} {2015})}\BibitemShut
  {NoStop}%
\bibitem [{\citenamefont {Hou}\ \emph {et~al.}()\citenamefont {Hou},
  \citenamefont {Li},\ and\ \citenamefont {Long}}]{Hou13}%
  \BibitemOpen
  \bibfield  {author} {\bibinfo {author} {\bibfnamefont {S.-Y.}\ \bibnamefont
  {Hou}}, \bibinfo {author} {\bibfnamefont {H.}~\bibnamefont {Li}}, \ and\
  \bibinfo {author} {\bibfnamefont {G.-L.}\ \bibnamefont {Long}},\ }\href
  {http://arxiv.org/pdf/1410.3940v1.pdf} {\bibinfo  {journal}
  {arXiv:1410.3940v1}\ }\BibitemShut {NoStop}%
\bibitem [{\citenamefont {Maletinsky}\ \emph {et~al.}(2012)\citenamefont
  {Maletinsky}, \citenamefont {Hong}, \citenamefont {Grinolds}, \citenamefont
  {Hausmann}, \citenamefont {Lukin}, \citenamefont {Walsworth}, \citenamefont
  {Loncar},\ and\ \citenamefont {Yacoby}}]{Maletinsky12}%
  \BibitemOpen
\bibfield  {journal} {  }\bibfield  {author} {\bibinfo {author} {\bibfnamefont
  {P.}~\bibnamefont {Maletinsky}}, \bibinfo {author} {\bibfnamefont
  {S.}~\bibnamefont {Hong}}, \bibinfo {author} {\bibfnamefont {M.~S.}\
  \bibnamefont {Grinolds}}, \bibinfo {author} {\bibfnamefont {B.}~\bibnamefont
  {Hausmann}}, \bibinfo {author} {\bibfnamefont {M.~D.}\ \bibnamefont {Lukin}},
  \bibinfo {author} {\bibfnamefont {R.~L.}\ \bibnamefont {Walsworth}}, \bibinfo
  {author} {\bibfnamefont {M.}~\bibnamefont {Loncar}}, \ and\ \bibinfo {author}
  {\bibfnamefont {A.}~\bibnamefont {Yacoby}},\ }\href
  {http://dx.doi.org/10.1038/nnano.2012.50} {\bibfield  {journal} {\bibinfo
  {journal} {Nat. Nanotech.}\ }\textbf {\bibinfo {volume} {7}},\ \bibinfo
  {pages} {320} (\bibinfo {year} {2012})}\BibitemShut {NoStop}%
\bibitem [{\citenamefont {Boss}\ \emph {et~al.}(2016)\citenamefont {Boss},
  \citenamefont {Chang}, \citenamefont {Armijo}, \citenamefont {Cujia},
  \citenamefont {Rosskopf}, \citenamefont {Maze},\ and\ \citenamefont
  {Degen}}]{Boss16}%
  \BibitemOpen
  \bibfield  {author} {\bibinfo {author} {\bibfnamefont {J.~M.}\ \bibnamefont
  {Boss}}, \bibinfo {author} {\bibfnamefont {K.}~\bibnamefont {Chang}},
  \bibinfo {author} {\bibfnamefont {J.}~\bibnamefont {Armijo}}, \bibinfo
  {author} {\bibfnamefont {K.}~\bibnamefont {Cujia}}, \bibinfo {author}
  {\bibfnamefont {T.}~\bibnamefont {Rosskopf}}, \bibinfo {author}
  {\bibfnamefont {J.~R.}\ \bibnamefont {Maze}}, \ and\ \bibinfo {author}
  {\bibfnamefont {C.~L.}\ \bibnamefont {Degen}},\ }\href
  {http://journals.aps.org/prl/pdf/10.1103/PhysRevLett.116.197601} {\bibfield
  {journal} {\bibinfo  {journal} {Phys. Rev. Lett.}\ }\textbf {\bibinfo
  {volume} {116}},\ \bibinfo {pages} {197601} (\bibinfo {year}
  {2016})}\BibitemShut {NoStop}%
\bibitem [{\citenamefont {Ljung}(1999)}]{Ljung99}%
  \BibitemOpen
  \bibfield  {author} {\bibinfo {author} {\bibfnamefont {L.}~\bibnamefont
  {Ljung}},\ }\href@noop {} {\emph {\bibinfo {title} {System Identification:
  Theory for the User, 2nd edition}}}\ (\bibinfo  {publisher} {Prentice-Hall,
  Upper Saddle River, NJ},\ \bibinfo {year} {1999})\BibitemShut {NoStop}%
\bibitem [{\citenamefont {Katayama}(2006)}]{Katayama06}%
  \BibitemOpen
  \bibfield  {author} {\bibinfo {author} {\bibfnamefont {T.}~\bibnamefont
  {Katayama}},\ }\href@noop {} {\emph {\bibinfo {title} {Subspace methods for
  system identification}}}\ (\bibinfo  {publisher} {Springer Science and
  Business Media, London},\ \bibinfo {year} {2006})\BibitemShut {NoStop}%
\bibitem [{\citenamefont {Caicedo}\ \emph {et~al.}(2014)\citenamefont
  {Caicedo}, \citenamefont {Dyke},\ and\ \citenamefont {Johnson}}]{Caicedo04}%
  \BibitemOpen
  \bibfield  {author} {\bibinfo {author} {\bibfnamefont {J.~M.}\ \bibnamefont
  {Caicedo}}, \bibinfo {author} {\bibfnamefont {S.~J.}\ \bibnamefont {Dyke}}, \
  and\ \bibinfo {author} {\bibfnamefont {E.~A.}\ \bibnamefont {Johnson}},\
  }\href
  {http://ascelibrary.org/doi/pdf/10.1061/(ASCE)0733-9399(2004)130:1(49)}
  {\bibfield  {journal} {\bibinfo  {journal} {J. Eng. Mech.}\ }\textbf
  {\bibinfo {volume} {130(1)}},\ \bibinfo {pages} {49} (\bibinfo {year}
  {2014})}\BibitemShut {NoStop}%
\bibitem [{\citenamefont {Moncayo}\ \emph {et~al.}(2010)\citenamefont
  {Moncayo}, \citenamefont {Marulanda},\ and\ \citenamefont
  {Thomson}}]{Moncayo10}%
  \BibitemOpen
  \bibfield  {author} {\bibinfo {author} {\bibfnamefont {H.}~\bibnamefont
  {Moncayo}}, \bibinfo {author} {\bibfnamefont {J.}~\bibnamefont {Marulanda}},
  \ and\ \bibinfo {author} {\bibfnamefont {P.}~\bibnamefont {Thomson}},\ }\href
  {http://ascelibrary.org/doi/pdf/10.1061/(ASCE)AS.1943-5525.0000011}
  {\bibfield  {journal} {\bibinfo  {journal} {J. Aerospace. Eng.}\ }\textbf
  {\bibinfo {volume} {23}},\ \bibinfo {pages} {99} (\bibinfo {year}
  {2010})}\BibitemShut {NoStop}%
\bibitem [{\citenamefont {Cox}\ \emph {et~al.}(2015)\citenamefont {Cox},
  \citenamefont {Little},\ and\ \citenamefont {O'Shea}}]{Cox15}%
  \BibitemOpen
  \bibfield  {author} {\bibinfo {author} {\bibfnamefont {D.~A.}\ \bibnamefont
  {Cox}}, \bibinfo {author} {\bibfnamefont {J.}~\bibnamefont {Little}}, \ and\
  \bibinfo {author} {\bibfnamefont {D.}~\bibnamefont {O'Shea}},\ }\href@noop {}
  {\emph {\bibinfo {title} {Ideals, Varieties, and Algorithms (An Introduction
  to Computational Algebraic Geometry and Commutative Algebra), 4th edition}}}\
  (\bibinfo  {publisher} {Springer, Cham},\ \bibinfo {year} {2015})\BibitemShut
  {NoStop}%
\bibitem [{\citenamefont {Cox}\ \emph {et~al.}(2005)\citenamefont {Cox},
  \citenamefont {Little},\ and\ \citenamefont {O'Shea}}]{Cox04}%
  \BibitemOpen
  \bibfield  {author} {\bibinfo {author} {\bibfnamefont {D.~A.}\ \bibnamefont
  {Cox}}, \bibinfo {author} {\bibfnamefont {J.}~\bibnamefont {Little}}, \ and\
  \bibinfo {author} {\bibfnamefont {D.}~\bibnamefont {O'Shea}},\ }\href@noop {}
  {\emph {\bibinfo {title} {Using Algebraic Geometry, 2nd edition}}}\ (\bibinfo
   {publisher} {Springer, New York},\ \bibinfo {year} {2005})\BibitemShut
  {NoStop}%
\bibitem [{\citenamefont {Fr\"oberg}(1997)}]{Froeberg97}%
  \BibitemOpen
  \bibfield  {author} {\bibinfo {author} {\bibfnamefont {R.}~\bibnamefont
  {Fr\"oberg}},\ }\href@noop {} {\emph {\bibinfo {title} {An Introduction to
  Gr\"obner basis}}}\ (\bibinfo  {publisher} {John Wiley \& Sons, New York},\
  \bibinfo {year} {1997})\BibitemShut {NoStop}%
\bibitem [{\citenamefont {Becker}\ and\ \citenamefont
  {Weispfenning}(1998)}]{Becker93}%
  \BibitemOpen
  \bibfield  {author} {\bibinfo {author} {\bibfnamefont {T.}~\bibnamefont
  {Becker}}\ and\ \bibinfo {author} {\bibfnamefont {V.}~\bibnamefont
  {Weispfenning}},\ }\href@noop {} {\emph {\bibinfo {title} {Gr\"obner basis: A
  Computational Approach to Commutative Algebra}}}\ (\bibinfo  {publisher}
  {Springer, New York},\ \bibinfo {year} {1998})\BibitemShut {NoStop}%
\bibitem [{\citenamefont {Buchberger}(2006)}]{Buchberger06}%
  \BibitemOpen
  \bibfield  {author} {\bibinfo {author} {\bibfnamefont {B.}~\bibnamefont
  {Buchberger}},\ }\href {\doibase 10.1016/j.jsc.2005.09.007} {\bibfield
  {journal} {\bibinfo  {journal} {J. Symb. Comput.}\ }\textbf {\bibinfo
  {volume} {41}},\ \bibinfo {pages} {475} (\bibinfo {year} {2006})}\BibitemShut
  {NoStop}%
\bibitem [{\citenamefont {Burgarth}\ and\ \citenamefont
  {Yuasa}(2012)}]{Burgarth12}%
  \BibitemOpen
  \bibfield  {author} {\bibinfo {author} {\bibfnamefont {D.}~\bibnamefont
  {Burgarth}}\ and\ \bibinfo {author} {\bibfnamefont {K.}~\bibnamefont
  {Yuasa}},\ }\href {\doibase 10.1103/PhysRevLett.108.080502} {\bibfield
  {journal} {\bibinfo  {journal} {Phys. Rev. Lett.}\ }\textbf {\bibinfo
  {volume} {108}},\ \bibinfo {pages} {080502} (\bibinfo {year}
  {2012})}\BibitemShut {NoStop}%
\bibitem [{\citenamefont {Burgarth}\ and\ \citenamefont
  {Yuasa}(2014)}]{Burgarth14}%
  \BibitemOpen
  \bibfield  {author} {\bibinfo {author} {\bibfnamefont {D.}~\bibnamefont
  {Burgarth}}\ and\ \bibinfo {author} {\bibfnamefont {K.}~\bibnamefont
  {Yuasa}},\ }\href {\doibase 10.1103/PhysRevA.89.030302} {\bibfield  {journal}
  {\bibinfo  {journal} {Phys. Rev. A}\ }\textbf {\bibinfo {volume} {89}},\
  \bibinfo {pages} {030302(R)} (\bibinfo {year} {2014})}\BibitemShut {NoStop}%
\bibitem [{\citenamefont {Gu\c{t}\v{a}}\ and\ \citenamefont
  {Yamamoto}(2016)}]{Guta13}%
  \BibitemOpen
  \bibfield  {author} {\bibinfo {author} {\bibfnamefont {M.}~\bibnamefont
  {Gu\c{t}\v{a}}}\ and\ \bibinfo {author} {\bibfnamefont {N.}~\bibnamefont
  {Yamamoto}},\ }\href {http://ieeexplore.ieee.org/abstract/document/7130587/}
  {\bibfield  {journal} {\bibinfo  {journal} {IEEE Trans. Autom. Control.}\
  }\textbf {\bibinfo {volume} {921}},\ \bibinfo {pages} {61} (\bibinfo {year}
  {2016})}\BibitemShut {NoStop}%
\bibitem [{\citenamefont {James}\ \emph {et~al.}(2008)\citenamefont {James},
  \citenamefont {Nurdin},\ and\ \citenamefont {Petersen}}]{James08}%
  \BibitemOpen
  \bibfield  {author} {\bibinfo {author} {\bibfnamefont {M.~R.}\ \bibnamefont
  {James}}, \bibinfo {author} {\bibfnamefont {H.~I.}\ \bibnamefont {Nurdin}}, \
  and\ \bibinfo {author} {\bibfnamefont {I.~R.}\ \bibnamefont {Petersen}},\
  }\href {\doibase 10.1109/TAC.2008.929378} {\bibfield  {journal} {\bibinfo
  {journal} {IEEE Trans Autom. Control.}\ }\textbf {\bibinfo {volume}
  {53(8)}},\ \bibinfo {pages} {1787} (\bibinfo {year} {2008})}\BibitemShut
  {NoStop}%
\bibitem [{\citenamefont {Yamamoto}(2014)}]{Yamamoto14}%
  \BibitemOpen
  \bibfield  {author} {\bibinfo {author} {\bibfnamefont {N.}~\bibnamefont
  {Yamamoto}},\ }\href {\doibase 10.1103/PhysRevX.4.041029} {\bibfield
  {journal} {\bibinfo  {journal} {Phys. Rev. X}\ }\textbf {\bibinfo {volume}
  {4}},\ \bibinfo {pages} {041029} (\bibinfo {year} {2014})}\BibitemShut
  {NoStop}%
\bibitem [{\citenamefont {Li}\ \emph {et~al.}(2009)\citenamefont {Li},
  \citenamefont {Ke},\ and\ \citenamefont {Ficek}}]{Li09}%
  \BibitemOpen
  \bibfield  {author} {\bibinfo {author} {\bibfnamefont {G.-X.}\ \bibnamefont
  {Li}}, \bibinfo {author} {\bibfnamefont {S.-S.}\ \bibnamefont {Ke}}, \ and\
  \bibinfo {author} {\bibfnamefont {Z.}~\bibnamefont {Ficek}},\ }\href
  {\doibase 10.1103/PhysRevA.79.033827} {\bibfield  {journal} {\bibinfo
  {journal} {Phys. Rev. A}\ }\textbf {\bibinfo {volume} {79}},\ \bibinfo
  {pages} {033827} (\bibinfo {year} {2009})}\BibitemShut {NoStop}%
\bibitem [{\citenamefont {Horn}\ and\ \citenamefont {Johnson}(2013)}]{Horn13}%
  \BibitemOpen
  \bibfield  {author} {\bibinfo {author} {\bibfnamefont {R.~A.}\ \bibnamefont
  {Horn}}\ and\ \bibinfo {author} {\bibfnamefont {C.~R.}\ \bibnamefont
  {Johnson}},\ }\href@noop {} {\emph {\bibinfo {title} {Matrix Analysis, 2nd
  edition}}}\ (\bibinfo  {publisher} {Cambridge University Press, New York},\
  \bibinfo {year} {2013})\BibitemShut {NoStop}%
\bibitem [{\citenamefont {Shannon}(1949)}]{Shannon49}%
  \BibitemOpen
  \bibfield  {author} {\bibinfo {author} {\bibfnamefont {C.~E.}\ \bibnamefont
  {Shannon}},\ }\href {\doibase 10.1109/JRPROC.1949.232969} {\bibfield
  {journal} {\bibinfo  {journal} {Proc. IRE 37.1 (1949): 10-21.}\ }\textbf
  {\bibinfo {volume} {37(1)}},\ \bibinfo {pages} {10} (\bibinfo {year}
  {1949})}\BibitemShut {NoStop}%
\bibitem [{\citenamefont {Bose}(2003)}]{Bose03}%
  \BibitemOpen
  \bibfield  {author} {\bibinfo {author} {\bibfnamefont {S.}~\bibnamefont
  {Bose}},\ }\href {\doibase 10.1103/PhysRevLett.91.207901} {\bibfield
  {journal} {\bibinfo  {journal} {Phys. Rev. Lett.}\ }\textbf {\bibinfo
  {volume} {91}},\ \bibinfo {eid} {207901} (\bibinfo {year}
  {2003})}\BibitemShut {NoStop}%
\bibitem [{\citenamefont {Franco}\ \emph {et~al.}(2008)\citenamefont {Franco},
  \citenamefont {Paternostro},\ and\ \citenamefont {Kim}}]{Franco08}%
  \BibitemOpen
  \bibfield  {author} {\bibinfo {author} {\bibfnamefont {C.~D.}\ \bibnamefont
  {Franco}}, \bibinfo {author} {\bibfnamefont {M.}~\bibnamefont {Paternostro}},
  \ and\ \bibinfo {author} {\bibfnamefont {M.~S.}\ \bibnamefont {Kim}},\ }\href
  {\doibase 10.1103/PhysRevLett.101.230502} {\bibfield  {journal} {\bibinfo
  {journal} {Phys. Rev. Lett.}\ }\textbf {\bibinfo {volume} {101}},\ \bibinfo
  {pages} {230502} (\bibinfo {year} {2008})}\BibitemShut {NoStop}%
\bibitem [{\citenamefont {Cappellaro}\ \emph {et~al.}(2011)\citenamefont
  {Cappellaro}, \citenamefont {Viola},\ and\ \citenamefont
  {Ramanathan}}]{Cappellaro11}%
  \BibitemOpen
  \bibfield  {author} {\bibinfo {author} {\bibfnamefont {P.}~\bibnamefont
  {Cappellaro}}, \bibinfo {author} {\bibfnamefont {L.}~\bibnamefont {Viola}}, \
  and\ \bibinfo {author} {\bibfnamefont {C.}~\bibnamefont {Ramanathan}},\
  }\href {\doibase 10.1103/PhysRevA.83.032304} {\bibfield  {journal} {\bibinfo
  {journal} {Phys. Rev. A}\ }\textbf {\bibinfo {volume} {83}},\ \bibinfo
  {pages} {032304} (\bibinfo {year} {2011})}\BibitemShut {NoStop}%
\bibitem [{\citenamefont {Ramanathan}\ \emph {et~al.}(2011)\citenamefont
  {Ramanathan}, \citenamefont {Cappellaro}, \citenamefont {Viola},\ and\
  \citenamefont {Cory}}]{Ramanathan11}%
  \BibitemOpen
  \bibfield  {author} {\bibinfo {author} {\bibfnamefont {C.}~\bibnamefont
  {Ramanathan}}, \bibinfo {author} {\bibfnamefont {P.}~\bibnamefont
  {Cappellaro}}, \bibinfo {author} {\bibfnamefont {L.}~\bibnamefont {Viola}}, \
  and\ \bibinfo {author} {\bibfnamefont {D.~G.}\ \bibnamefont {Cory}},\ }\href
  {\doibase 10.1088/1367-2630/13/10/103015} {\bibfield  {journal} {\bibinfo
  {journal} {New J. Phys.}\ }\textbf {\bibinfo {volume} {13}},\ \bibinfo
  {pages} {103015} (\bibinfo {year} {2011})}\BibitemShut {NoStop}%
\bibitem [{\citenamefont {Haeberlen}(1976)}]{Haeberlen76}%
  \BibitemOpen
  \bibfield  {author} {\bibinfo {author} {\bibfnamefont {U.}~\bibnamefont
  {Haeberlen}},\ }\href@noop {} {\emph {\bibinfo {title} {High Resolution NMR
  in Solids: Selective Averaging}}}\ (\bibinfo  {publisher} {Academic Press
  Inc., New York},\ \bibinfo {year} {1976})\BibitemShut {NoStop}%
\bibitem [{\citenamefont {Magnus}(1954)}]{Magnus54}%
  \BibitemOpen
  \bibfield  {author} {\bibinfo {author} {\bibfnamefont {W.}~\bibnamefont
  {Magnus}},\ }\href@noop {} {\bibfield  {journal} {\bibinfo  {journal}
  {Communications on Pure and Applied Mathematics}\ }\textbf {\bibinfo {volume}
  {7}},\ \bibinfo {pages} {649} (\bibinfo {year} {1954})}\BibitemShut {NoStop}%
\bibitem [{\citenamefont {Markovsky}(2012)}]{Markovsky12}%
  \BibitemOpen
  \bibfield  {author} {\bibinfo {author} {\bibfnamefont {I.}~\bibnamefont
  {Markovsky}},\ }\href@noop {} {\emph {\bibinfo {title} {Low rank
  approximation: algorithms, implementation, applications}}}\ (\bibinfo
  {publisher} {Springer Communications and Control Engineering, London; New
  York},\ \bibinfo {year} {2012})\BibitemShut {NoStop}%
\bibitem [{\citenamefont {Lakshman}\ and\ \citenamefont
  {Lazard}(1991)}]{Lakshman91}%
  \BibitemOpen
  \bibfield  {author} {\bibinfo {author} {\bibfnamefont {Y.~N.}\ \bibnamefont
  {Lakshman}}\ and\ \bibinfo {author} {\bibfnamefont {D.}~\bibnamefont
  {Lazard}},\ }\href
  {http://link.springer.com/chapter/10.1007/978-1-4612-0441-1_14} {\bibfield
  {journal} {\bibinfo  {journal} {In Effective methods in algebraic geometry,
  Progress in Mathematics (Birkh\"{a}user, Boston)}\ }\textbf {\bibinfo
  {volume} {94}},\ \bibinfo {pages} {217} (\bibinfo {year} {1991})}\BibitemShut
  {NoStop}%
\bibitem [{\citenamefont {Gritzmann}\ and\ \citenamefont
  {Sturmfels}(1993)}]{Gritzmann93}%
  \BibitemOpen
  \bibfield  {author} {\bibinfo {author} {\bibfnamefont {P.}~\bibnamefont
  {Gritzmann}}\ and\ \bibinfo {author} {\bibfnamefont {B.}~\bibnamefont
  {Sturmfels}},\ }\href {http://epubs.siam.org/doi/pdf/10.1137/0406019}
  {\bibfield  {journal} {\bibinfo  {journal} {SIAM J. Discrete Math.}\ }\textbf
  {\bibinfo {volume} {6}},\ \bibinfo {pages} {246} (\bibinfo {year}
  {1993})}\BibitemShut {NoStop}%
\bibitem [{\citenamefont {Caboara}\ and\ \citenamefont
  {Perry}(2014)}]{Caboara14}%
  \BibitemOpen
  \bibfield  {author} {\bibinfo {author} {\bibfnamefont {M.}~\bibnamefont
  {Caboara}}\ and\ \bibinfo {author} {\bibfnamefont {J.}~\bibnamefont
  {Perry}},\ }\href
  {http://link.springer.com/article/10.1007%2Fs00200-014-0216-5} {\bibfield
  {journal} {\bibinfo  {journal} {Appl. Algebra Eng. Comm. Comput.}\ }\textbf
  {\bibinfo {volume} {25}},\ \bibinfo {pages} {99} (\bibinfo {year}
  {2014})}\BibitemShut {NoStop}%
\end{thebibliography}%

\end{document}